\documentclass[final,5p,times,twocolumn]{elsarticle}
\biboptions{numbers,sort&compress,square}

\usepackage{amssymb}
\usepackage{amsmath}
\usepackage{csquotes} 
\usepackage[hidelinks]{hyperref}
\usepackage{upgreek} 
\usepackage[version=4]{mhchem} 
\usepackage{siunitx}
\usepackage{dblfloatfix}  
\usepackage{placeins}  
\usepackage{caption} 

\usepackage{comment}

\usepackage{xcolor}

\usepackage{booktabs}
\usepackage{siunitx}

\journal{Carbon}

\begin{document}


\begin{frontmatter}



\title{Carbon black and hydrogen production from methane pyrolysis: measured and modeled insights from integrated gas and particle diagnostics in shock tubes}


\author[a]{Gibson Clark}
\author[b]{Mohammad Adib}
\author[a]{Chengze Li}
\author[a]{Taylor M. Rault}
\author[a]{\\Jesse W. Streicher}
\author[c]{Enoch Dames}
\author[b]{M. Reza Kholghy}
\author[a]{Ronald K. Hanson}

\affiliation[a]{organization={Department of Mechanical Engineering, Stanford University},
            addressline={452 Escondido Mall, Bldg 520}, 
            city={Stanford},
            postcode={94305}, 
            state={CA},
            country={USA}}

\affiliation[b]{organization={Department of Mechanical Engineering, Carleton University},
            addressline={1125 Colonel By Drive}, 
            city={Ottawa},
            postcode={K1S 5B6}, 
            state={ON},
            country={Canada}}

\affiliation[c]{organization={Monolith Materials},
            city={San Carlos},
            postcode={94070}, 
            state={CA},
            country={USA}}

\begin{abstract}
Methane (\ce{CH4}) pyrolysis is a promising route to co-produce hydrogen (\ce{H2}) and carbon black (CB) while avoiding emissions associated with steam-methane reforming and furnace black processes. Model development of pyrolytic CB synthesis requires experimental observations of concurrent gas chemistry, particulate formation, and morphology. This work presents a combined experimental and modeling study of \ce{CH4} pyrolysis behind reflected shock waves in 5\% \ce{CH4}/Argon mixtures at post-reflected-shock temperatures ($T_5$) of 1850--2450~K and $P_5 \approx 4.5 \pm 0.4$~atm. Laser absorption diagnostics quantified \ce{CH4}, \ce{C2H4}, and \ce{C2H2} mole fractions, while multiwavelength extinction (633 and 1064~nm) resolved time-dependent particle formation and the temperature-dependent evolution of optical maturity. Simulations reproduce small-molecule speciation well, but large variations in predicted polycyclic aromatic hydrocarbons (PAHs) persist among models. Coupled gas--particle simulations capture accurate volume fraction ($f_\mathrm{v}$) trends and the influence of gas dynamics but underpredict induction times at high $T_5$. Samples collected at the shock tube endwall were analyzed by transmission electron microscopy (TEM) to quantify primary particle size distributions and nanostructure arrangement. Image segmentation and manual measurements showed reduced primary particle size growth ($d_\mathrm{p}$) with increasing $T_5$, while graphitic nanostructure generally increased. This study provides an integrated benchmark for improving models of CB and \ce{H2} production from \ce{CH4} pyrolysis by constraining gas-phase kinetics, PAH-driven inception, particle dynamics, and particle maturity. The results highlight that accurate partitioning of mass between particle number and particle size is an important constraint for further model development.

\end{abstract}


\end{frontmatter}



\section{Introduction}
\label{intro}

\par Carbon black (CB) and hydrogen gas (H$_2$) are high-value commodities with wide-ranging industrial applications. CB, used extensively as a reinforcing agent in rubbers and tires, has an annual global production of $\sim8$ Mt \cite{fanPresentFutureCarbon2020a}. However, current CB production relies on partial combustion which suffers from low yields and significant inefficiencies in the form of CO$_2$ emissions \cite{Donnet1993_CBtextbook}. In parallel, H$_2$ is used in many industrial processes. Global H$_2$ demand is nearly 100 Mt annually \cite{IEA_h2_2025}. At present, over 95\% of H$_2$ production is derived from steam methane reforming (SMR), which accounts for $\sim$2\% of global greenhouse gas (GHG) emissions \cite{ishaqReviewHydrogenProduction2022}.

\par The pyrolysis of methane (CH$_4$) offers a pathway to co-produce CB and H$_2$ without direct GHG emissions \cite{sanchez-bastardoMethanePyrolysisZeroEmission2021}. Early commercial approaches combine existing natural gas infrastructure with reactor innovations like thermal plasma pyrolysis and show potential for low emissions, high purity, and nearly 100\% product yields \cite{fulcheriEnergyefficientPlasmaMethane2023}. However, achieving high-purity and high-grade H$_2$ and CB products remains difficult. It is a formidable challenge to control the formation of carbon allotropes (CB, graphite, fullerenes, nanotubes) with specific properties, (e.g., primary particle diameter, $d_\mathrm{p}$; specific surface area, SSA), when inception, surface growth, coalescence, and agglomeration processes and the resulting internal nanostructure depend sensitively on feedstock, chemical pathways, local gas temperature, pressure, and occur on sub-millisecond timescales \cite{wangFormationNascentSoot2011, vanderwalSootOxidation2003,violiRelativeRolesAcetylene2005,thomsonModelingSootFormation2023}.

\par CB formation from high-temperature CH$_4$ pyrolysis proceeds first by small radical production (H, CH$_\mathrm{x}$) and fuel conversion into  gas intermediates. Important intermediates such as acetylene (C$_2$H$_2$), which is abundant in high temperature hydrocarbon pyrolysis \cite{frenklachSootFormationBinary1988, frenklachMechanismSootNucleation2020}, and resonantly stabilized radicals (RSRs), including propargyl (C$_3$H$_3$) \cite{rajC3H32014}, serve as precursors to the formation of polycyclic aromatic hydrocarbons (PAHs) \cite{richterHoward2000, hamadiThesis2021, thomsonModelingSootFormation2023}. The buildup of PAH species provides a pool of aromatic molecules that combine to form incipient carbonaceous particles \cite{martinSootInceptionCarbonaceous2022, frenklachMechanismSootNucleation2020, lieskePortraitsSootMolecules2023}; there is building experimental evidence that the dimerization of PAHs is a crucial step in this formation process \cite{mercierDimersPolycyclicAromatic2019, faccinettoEvidenceFormationDimers2020}. Once formed, particle mass is believed to increase via surface growth, which includes both the adsorption of PAHs and by the Hydrogen-Abstraction-Carbon-Addition mechanism (HACA) \cite{frenklachDetailedModelingSoot1991}, more generally referred to as H-Activated-Carbon-Addition due to aromatic site activation by both H addition and H abstraction \cite{frenklachMigrationMechanismAromaticedge2005, frenklachMechanismSootNucleation2020}. As particles form they may also undergo coalescence (merging) into larger, spherical, primary particles. In a typical reactor, as they age, these primary particles undergo agglomeration (loosely bound grouping), aggregation (tightly bound grouping), graphitization (dehydrogenation and increasing nanostructural order), and finally---under certain conditions---oxidation and fragmentation \cite{richterHoward2000, michelsenProbingSootFormation2017, thomsonModelingSootFormation2023}. An excellent summary of particle aging processes is given by Michelsen et al. (2020) \cite{michelsenReviewTerminologyUsed2020}. Numerical modeling of this process is complicated by uncertainty in gas species concentrations, the reversibility of PAH kinetics, debated inception mechanisms, heterogeneous gas--particle reactions, and evolving particle dynamics \cite{wangFormationNascentSoot2011, frenklachMechanismSootNucleation2020, martinSootInceptionCarbonaceous2022, thomsonModelingSootFormation2023}. The complex interactions of these processes necessitate reliable measurements of CB formation rates, yields, and morphology at reactor-relevant temperatures to benchmark model development. Unlike flames (oxidative), flow reactors (longer timescales), or plasmas (diagnostically limited), the shock tube provides a uniquely controlled platform to optically probe high-temperature CH$_4$ pyrolysis with multiple simultaneous diagnostics at shock-heated, near-isobaric conditions subject to limited diffusion and mixing \cite{ereminFormationCarbonNanoparticles2012}.

\par Accurate depictions of gas chemistry are required to understand and model CB formation from CH$_4$ pyrolysis. Considerable effort has gone into measuring fuel pyrolysis and combustion chemistry using laser absorption spectroscopy (LAS) in shock tubes \cite{hanson_LAS_shockTubes_2014}. LAS can provide a non-intrusive, time-resolved, and species-specific means of gas multi-speciation \cite{pinkowskiMultiwavelengthSpeciationFramework2019}. Significant progress has led to the development and validation of kinetic models such as FFCM-2 \cite{zhangFoundationalFuelChemistry2023} and NUIG 1.3 \cite{panigrahyWhenHydrogenSlower2023} which are designed to accurately describe the kinetics of hydrocarbons up to C$_4$. During particle synthesis, however, carbon flux proceeds through aromatic molecules of numerous rings \cite{lieskePortraitsSootMolecules2023}. As a result, accurate modeling of pyrolytic CB synthesis requires kinetic schemes that extend to C$_{16}$ or larger PAHs such as works by \cite{wangDetailedKineticModeling1997, appelKineticModelingSoot2000, blanquartChemicalMechanismHigh2009, wangPAHGrowthMechanism2013, hamadiThesis2021}. The kinetics of CH$_4$ thermal decomposition have been thoroughly investigated under dilute conditions with the main products being C$_2$H$_2$ and C$_2$H$_4$ \cite{hidakaShocktubeModelingStudy1999, fauMethanePyrolysisLiterature2013, gueretMethanePyrolysisThermodynamics1997} and the rates of key reactions, such as the highly endothermic the initiation step CH$_4$+M $\rightarrow$ CH$_3$ + H + M (R1), are very well known \cite{wangImprovedShockTube2016, shaoShockinducedIgnitionPyrolysis2020}. At higher fuel loadings, however, recent studies suggest that measurement-model discrepancies remain, even in small molecule chemistry \cite{nativelShocktubeStudyMethane2019, ferrisExperimentalNumericalInvestigation2024}. Overall, despite extensive literature, a significant lack of knowledge about the pathways leading to particle inception and rates of carbon flux from CH$_4$ into PAHs and ultimately solid carbonaceous particles persist \cite{agafonovUnifiedKineticModel2016, thomsonModelingSootFormation2023}.

\par The study of particle formation in shock tubes has been thoroughly advanced with optical diagnostics such as light extinction, scattering, and laser-induced incandescence (LII) \cite{ereminFormationCarbonNanoparticles2012, michelsenProbingSootFormation2017, kholghyComparisonMultipleDiagnostic2017, ereminPassiveActiveLaser2025}. Light extinction can provide crucial phenomenological constraints for nanoparticle growth models. It is a sensitive means of detecting the formation of condensed-phase material due to the high refractive index of particles relative to the gas medium \cite{ereminFormationCarbonNanoparticles2012}. The combined absorption and scattering of transmitted light are determined by the complex refractive index, $m$, Eq.~(\ref{refractiveIndex}) in Section \ref{methods_LEx}. Provided the Rayleigh criterion is met (particles much smaller than the transmitted wavelength), absorption dominates, and parameters such as condensed phase volume fraction ($f_\mathrm{v}$), induction time $\tau_\mathrm{ind}$, global growth rates, and carbon yield ($\mathrm{CY}$), can be quantified through the absorption function, $E(m)$, as detailed in Section \ref{methods_LEx}, Eq.~(\ref{absorptionFunction}). $E(m)$ arises from electromagnetic theory as a fundamental property of light absorption \cite{bohren1983} and represents an effective absorption cross section in Rayleigh extinction experiments. $E(m)$ has been studied both experimentally with optical techniques including LII and light extinction in flames \cite{liuReviewRecentLiterature2020} and theoretically with calculations based on Rayleigh-Debye-Gans theory (RDG) \cite{dobbinsAbsorptionScatteringLight1991} and updated with discrete dipole approximation (DDA) to consider the electromagnetic coupling of primary particle monomers within an aggregate \cite{yonRadiativePropertiesSoot2015}.

\par Reported $E(m)$ values range from $< 0.1$ for incipient particles \cite{ereminSizeDependenceComplex2011} to 0.4 for mature carbon particles \cite{liuReviewRecentLiterature2020}. A fixed $E(m)$ is commonly applied in a given analysis and a recent review by Liu et al. \cite{liuReviewRecentLiterature2020} suggested that $E(m) \ge 0.32$ for mature CB. Crucially, however, short residence times, significant inception flux, and rapidly evolving particle size and maturity introduce error in an assumption of a fixed $E(m)$ due to the dependence of $m$ on particle composition \cite{bescondSootOpticalProperties2016, kelesidisImpactOrganicCarbon2021}, nanostructure \cite{gurentsovEffectSizeStructure2022}, and size \cite{wanQuantumConfinementSize2021, kelesidisDeterminationVolumeFraction2021}. For example, Wan et al., \cite{wanQuantumConfinementSize2021} showed that a model based on quantum confinement and amorphous semiconductor theory of monodisperse particles predicts a 3$\times$ increase in the absorption aspect of refractive index, $k$, at 633 nm as particle volume median diameter grows from 5 to 15 nm. Other measurements by Eremin et al. using LII report $E(m)$ increasing from 0.05 to 0.25 as particles evolve and grow up to 20 nm in C$_2$H$_2$ pyrolysis \cite{ereminSizeDependenceComplex2011}.

\par Importantly, the complex refractive index $m$, and therefore $E(m)$, exhibit a strong spectral dependence from the UV to near infrared (NIR) related to the particle’s physical characteristics \cite{michelsenProbingSootFormation2017, miglioriniInvestigationOpticalProperties2011, hagenCarbonNanostructureReactivity2021,ereminSpectralDependenceAbsorbance2024}. A method connecting particle H/C ratio and nanostructure with spectrally-dependent optical properties was proposed by Bescond et al. in 2016, in which optical constants from three reference aerosols representing organic, amorphous, and ordered (graphitic) particulate were combined using a linear mixing rule \cite{bescondSootOpticalProperties2016}. At shorter wavelengths ($< 685 - 700$ nm), organic PAH molecules are thought to exhibit high-temperature absorption features that can contribute to light attenuation \cite{miglioriniInvestigationOpticalProperties2011, ereminSpectralDependenceAbsorbance2024}. Short wavelength extinction is also thought to be more sensitive to volatile organic materials that comprise incipient particles and condense on particle surfaces \cite{michelsenProbingSootFormation2017}. However, at longer wavelengths, extinction shows a more selective sensitivity to more mature, graphitic particles \cite{yonRevealingSootMaturity2021} due to their reduced optical band gap that comes with increased order and $\mathrm{sp}^2$ hybridization \cite{russoOpticalBandGap2020,hagenCarbonNanostructureReactivity2021}.  Multiwavelength extinction from visible to NIR can therefore help bound uncertainty in optical properties while providing insight on the progression of size and composition from a particle with a high organic carbon content to one that is highly graphitic (i.e. particle maturity) \cite{michelsenProbingSootFormation2017, limMeasurementOrganicCarbon2019}.

\par Morphological data is crucial for model development to control the grade and properties of the synthesized carbon product. Transmission electron microscopy (TEM) and high-resolution TEM (HRTEM) permit observations of micron-scale aggregate morphology, nanometer-scale primary particle sizes ($d_\mathrm{p}$), and angstrom-scale internal nanostructure to inform fundamental understandings of a particle’s physical nature and guide modeling efforts \cite{kholghyCoreShellInternal2016, baldelliDeterminingSootMaturity2020}. While extensively studied in flames, the literature on detailed morphology of particulate formed in thermal pyrolysis is comparatively sparse. Baurle et al. (1994) reported narrow lognormal distributions of primary particles centered between 20--30 nm and found no influence of pressure or fuel H/C ratio in benzene (\ce{C6H6}) and n-hexane pyrolysis \cite{bauerleSootFormationElevated1994}. Studies of heavy hydrocarbons by Douce et al. (2000) \cite{douceSootFormationHeavy2000} and toluene by Mathieu (2007) \cite{mathieuCharacterizationAdsorbedSpecies2007} have since measured a decrease in primary particle size with increasing post-shock temperature ($T_\mathrm{5}$) in the range of 1500--2200 K, but the search for a mechanistic consensus remains. HRTEM-derived nanostructure enable further insight into CB evolution by quantifying how lattice fringes organize, curve, and stack within the particles. The analysis methodology of the field has therefore evolved from predominantly qualitative descriptors toward automated fringe-extraction that transforms raw micrographs into nonstructural maps, enabling the derivation of statistically meaningful distributions of shape metrics \cite{tothNanostructureQuantificationTurbostratic2021}. In this framework, three descriptors are adopted: the fringe length $L_\mathrm{f}$, defined as the end-to-end distance measured along an individual fringe; the fringe tortuosity $\tau$, defined as the ratio of that fringe length to the straight-line distance between the fringe endpoints; and the fringe spacing $d_\mathrm{f}$, which characterizes the average separation between two adjacent carbon fringes in the same stack \cite{singhSootDifferentiationLaser2019}. Synthesis studies based on both fuel pyrolysis and oxidation indicate that these quantities depend not only on $T_5$, or peak temperature during the processes, but on reactant composition, residence time, and thermal history, such that particles generated at different conditions exhibit distinct internal textures and inferred maturity  \cite{vanderwalSootNanostructureDependence2004}. In shock-tube pyrolysis, Gurentsov et al.\ reported decreasing internal spacing with particle growth and increasing order as $T_5$ increases across small-hydrocarbon systems \cite{gurentsovEffectSizeStructure2022}, whereas toluene pyrolysis studies by Mathieu et al. suggests a non-monotonic ordering, with the longest fringes occurring at intermediate temperatures and multiple particle types present at the highest $T_5$ (near 2135\,K) \cite{mathieuCharacterizationAdsorbedSpecies2007}. These results motivate analysis of $L_\mathrm{f}$, $\tau$, and $d_\mathrm{f}$ alongside operating conditions to interpret CB maturity and structural diversity from CH$_4$ pyrolysis.
\par Complementary time-resolved and morphological datasets provide the strongest constraints for models to capture not only formation rates and yields, but how particle structure changes with process variables \cite{carbone1ProbingGastoparticleTransition2017, carbone2ProbingGastoparticleTransition2017}, yet few experiments have reported such results in shock tubes. Combined C$_2$H$_2$ and extinction time-histories were recently reported in C$_6$H$_6$ and C$_2$H$_4$ pyrolysis behind shock waves \cite{kcSimultaneousMeasurementsAcetylene2017} and Agafonov et al. (2016) developed an extended kinetic model benchmarked against gas-species and soot yield measurements of numerous aliphatic and aromatic fuel pyrolysis \cite{agafonovUnifiedKineticModel2016}. To the authors’ knowledge, an integrated study of gas speciation, time-resolved particle formation, and detailed morphology to provide a more complete picture of CB synthesis from CH$_4$ pyrolysis has yet to be reported. For this reason, the goals of this combined experimental and modeling study are threefold:
\begin{enumerate}
    \item Evaluate kinetic model agreement with gas-phase chemistry under \ce{CH4} pyrolysis conditions for the purpose of describing the detailed chemistry in a numerical model of carbon particle formation 
    \item  Bring maturity-specific insights to induction time, growth rates, and yield during high-temperature particle synthesis via multiwavelength extinction
    \item Leverage particle sampling and TEM-derived morphology and nanostructure to bridge between yield-focused and structure-focused modeling frameworks and better interpret the time-resolved data and model predictions
\end{enumerate}

\par To address these goals, we performed simultaneous measurements of CH$_4$, C$_2$H$_4$, C$_2$H$_2$, and condensed-phase particulate during the pyrolysis of 5\% CH$_4$ in Ar at post-reflected shock pressures ($P_\mathrm{5}$) near 4.5 atm and post-reflected shock temperatures ($T_\mathrm{5}$) between 1850--2450 K, while CB samples were collected for TEM imaging and analysis. Omnisoot, a computational modeling platform coupling detailed gas chemistry, particle dynamics, and inception and growth models has recently been developed and applied throughout this work \cite{adibOmnisootProcessDesign2026}. While time-histories enable direct assessment of Omnisoot's ability to capture the carbon flux from the gas phase into particles, $d_\mathrm{p}$ measurements from TEM analysis highlight crucial areas to target for improving model fidelity. By linking speciation, extinction, and morphology in a single analysis, we take a step towards more predictive models of CB yield, structure, and maturity to improve design tools in advanced pyrolytic synthesis processes.

 
\subsection*{A note on terminology}  
Because carbonaceous particle formation lies at the interface of multiple disciplines, terminology is often inconsistent. To promote clarity, we adopt the definitions and guidelines proposed by Michelsen et al. (2020) \cite{michelsenReviewTerminologyUsed2020}.

\section{Methods}
\label{methods}

\subsection{Shock tube experimental facility}
\label{methods_shockTube}

\par
\begin{figure}[h]
    \centering
    \includegraphics[width=1.0\linewidth]{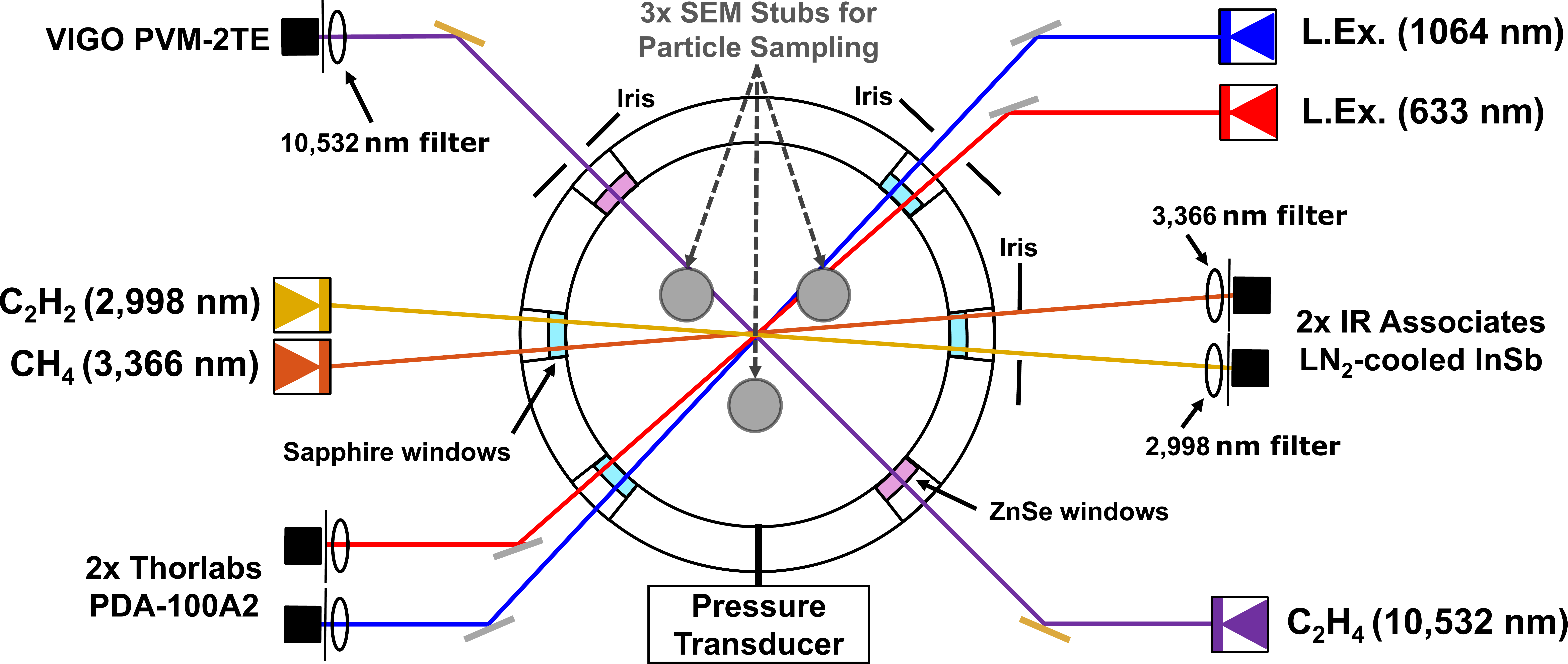}
    \caption{Schematic of gas speciation, particle formation, and particle sampling diagnostic integration for shock-tube pyrolysis experiments.}
    \label{fig:fig1}
\end{figure}

 Experiments were conducted in the Stanford Flexible Applications Shock Tube (FAST) with a 3.35 m driver section, 9.70 m driven section, and a 14.12 cm inner diameter. Driver inserts were used to extend the duration of near-constant post-reflected-shock pressures at the measurement plane which was located 9 mm from the endwall. Five pressure transducers (PTs, PCB 113A26) measured incident shock velocities to compute $T_5$ and $P_5$ from normal shock relations using a custom code assuming frozen chemistry and complete vibrational relaxation \cite{campbellFROSH2017}. A sidewall PT (Kistler 603B1) and three pairs of optical ports enabled pressure and laser attenuation signals to be recorded at 125 MHz on two Pico Technology 5444D oscilloscopes. A total of five lasers were aligned through the shock tube to measure the evolution of CH$_4$, C$_2$H$_4$, C$_2$H$_2$, and carbon particle formation (Fig.~\ref{fig:fig1}), while the synthesized CB was thermophoretically collected on scanning electron microscopy (SEM) stubs mounted in a custom shock tube endwall for \textit{ex situ} analysis. Irises and narrow band-pass filters spatially and spectrally filtered broadband emission from the hot gases to ensure high-quality transmission signals were recorded on all photodetectors. Gases were supplied by Linde Gas \& Equipment, including CH$_4$ (99.9\% purity) and Ar (99.999\% purity). Mixtures of 5\% CH$_4$ in argon were prepared monometrically to 3000 Torr in a mixing tank (0.01 Torr ultimate pressure) and mixed for at least 30 minutes by a magnetic stirrer before an experiment.

\subsection{Laser absorption diagnostics}
\label{methods_LAS}

Gas speciation measurements of carbon flux used the Beer-Bouguer-Lambert relation between fractional light transmission $I_{\mathrm{t}}/I_{\mathrm{0}}$ and spectroscopic and thermodynamic properties of the absorbing molecule along a homogeneous line of sight:

\begin{equation}
    -\ln\left(\frac{I_{\mathrm{t}}}{I_{\mathrm{0}}}\right)_{\lambda} = \alpha_{\lambda} = \sum_i k_{i,\lambda}(T, P) \chi_i P L = \sum_i \sigma_{i,\lambda}(T, P) \chi_i n L
    \label{beer-lambert}   
\end{equation}

where $\alpha_{\lambda}$ is the net absorbance at wavelength $\lambda$, $k_{i,\lambda}$ is the absorption coefficient for species $i$, $P$ is the total pressure, $\chi_i$ is the species mole fraction, and $L$ is the path length (14.12 cm). For blended spectra of larger molecules, absorption cross-section $\sigma_{i,\lambda}$ and total number density $n$ can be related the same way. Summation over $i$ species accounts for the potential of interfering absorbance, which can be corrected using known values of $k_{i,\lambda}$ or $\sigma_{i,\lambda}$.

Absorbance time-histories of $\mathrm{C}_2\mathrm{H}_4$ were recorded on a Vigo Photonics PVM-2TE-10.6-3x3-TO8 IR detector by measuring the P14 transition of the $\mathrm{CO}_2$ 0,0,1$\rightarrow$1,0,0 vibrational band at 10.532 $\upmu\mathrm{m}$ (Access Laser Co. L4 series $\mathrm{CO}_2$-gas laser). Absorption cross sections were characterized by Ren et al. \cite{renIRLaserAbsorption2012} and updated in \cite{pinkowskiMultiwavelengthSpeciationFramework2019}.


$\mathrm{C}_2\mathrm{H}_2$ was measured at 2.998 $\upmu\mathrm{m}$ by a NanoPlus DFB laser using a LN$_2$-cooled IR Associates IS-2.0 InSb detector. Absorption coefficients were first measured by Stranic et al. (2014) \cite{stranicLaserAbsorptionDiagnostic2014} and updated for this work: the high $\mathrm{CH}_4$ loadings used herein to promote CB synthesis create interfering absorption from $\mathrm{CH}_4$ near 2.998 $\upmu\mathrm{m}$ at early times and low $T_{5}$ (when $\mathrm{CH}_4$ concentrations are large). To infer accurate $\mathrm{C}_2\mathrm{H}_2$ mole fractions free of interference, a temperature-dependent interfering $k_{\mathrm{CH_4},\lambda}(T)$ for $\mathrm{CH}_4$ was measured at $\lambda=$ 2.998 $\upmu\mathrm{m}$ (Section \ref{SM:absorption_coefficients} in the Supplementary Materials, S.M.). The interference accounted for using this method resulted in corrections to the absolute measured mole fraction of $\mathrm{C}_2\mathrm{H}_2$ between 0.001 and 0.004 (2-10\% of the actual \ce{C2H2} concentration formed during the test-time). 

A new absorption diagnostic using a NanoPlus DFB diode laser was developed to measure $\mathrm{CH}_4$ via a cluster of C-H stretch vibrations in the P-branch of $\mathrm{CH}_4$ near 3366 nm. This wavelength reduces spectral interference from high-temperature $\mathrm{C}_2\mathrm{H}_4$ and $\mathrm{C}_2\mathrm{H}_2$ transitions near alternative \ce{CH4} diagnostics \cite{surHighsensitivityInterferencefreeDiagnostic2015, ferrisExperimentalNumericalInvestigation2024} and reduces sensitivity to temperature. Temperature-dependent $\mathrm{CH}_4$ absorption coefficients at 3366 nm were quantified behind reflected shock waves from 1500--2450 K and 1.0--5.0 atm and are reported in Section \ref{SM:absorption_coefficients} of the S.M. Laser light attenuation was measured on a LN$_2$-cooled IR Associates IS-2.0 InSb detector.

\begin{figure*}[t]
    \centering
    \includegraphics[width=\textwidth]{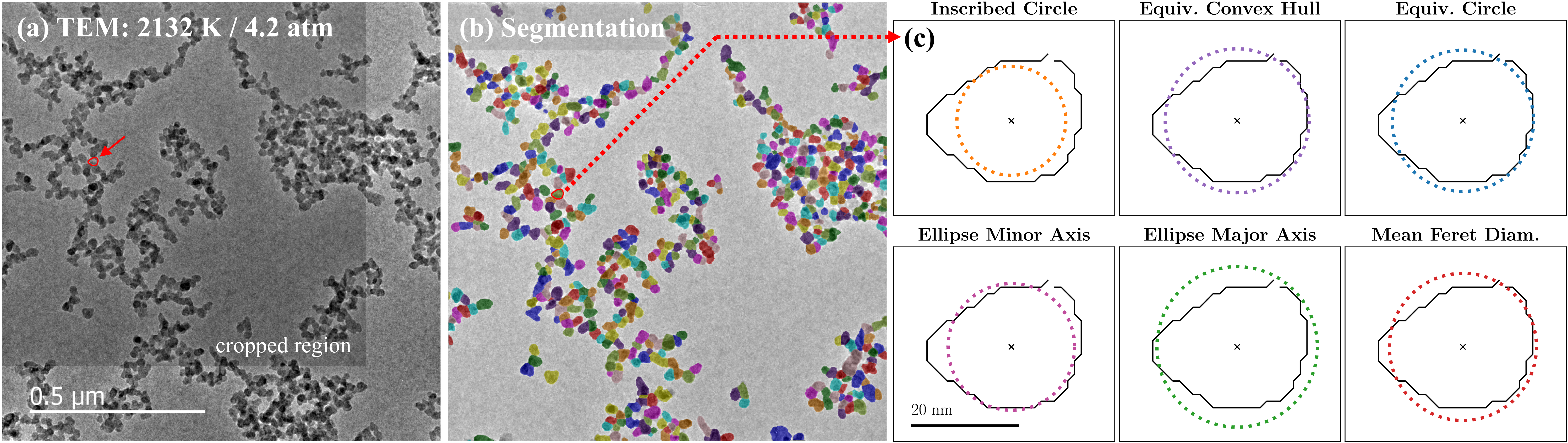}
    \caption{Segmentation of a representative image from 2132\,K, 4.2\,atm using Cellpose-SAM \cite{pachitariu_CellposeSAMS} and the effective diameter metrics employed in this study: (a) Raw TEM image, which was cropped to the dark region for automated analysis, with the example contour (339) highlighted in red, (b) Processed TEM segmentation mask showing 678 detected particle contours, (c) Contour 339 and its effective primary particle diameter described by six different size metrics.}
    \label{fig:f_cellposeMethods}
\end{figure*}

\subsection{Light extinction diagnostics}
\label{methods_LEx}

Laser light extinction at 633 nm (Coherent Inc. 70 mW OBIS 633 nm LX laser) and 1064 nm (Edmund Optics 50 mW LC turnkey laser) was used to resolve the formation of carbon particles in the shock tube. Laser light intensity was recorded on Thorlabs PDA-100A2 detectors. The complex refractive index ($m$) of a condensed phase characterizes the total attenuation of transmitted light via absorption and scattering where $n$ and $k$ reflect scattering and absorption contributions respectively:

\begin{equation}
    m = n - ik
    \label{refractiveIndex}
\end{equation}

Following the RDG framework in the Rayleigh limit of primary particle size $d_\mathrm{p}$ $\ll \lambda$, (confirmed here by TEM), absorption dominates light attenuation, and the extinction coefficient ($\varepsilon$) can be entirely described by the absorption function $E(m)$:

\begin{equation}
    E(m,\lambda) = \frac{\varepsilon \lambda}{6 \pi} = -h\times\mathrm{Im}\left\{ \frac{m_{\lambda}^2 - 1}{m_{\lambda}^2 + 2} \right\}
    \label{absorptionFunction}
\end{equation}

\noindent where $h$ is used to correct for the light absorption enhancement due to multiple light scattering between primary particles estimated from DDA \cite{yonExtensionRDGFAScattering2008, yonRadiativePropertiesSoot2015}. In classic RDG theory applied herein, $h = 1$, while recent work has approximated $h$  between 1 and 1.2 for similar morphology and wavelengths \cite{kelesidisDeterminationVolumeFraction2021}. The Beer-Bouguer-Lambert law can then be used to infer particle volume fractions ($f_{\mathrm{v}}$) $[\mathrm{m}_{\mathrm{prt}}^3/\mathrm{m}_\mathrm{{gas}}^3]$ from laser light extinction measurements:

\begin{equation}
    -\ln\left(\frac{I_{\mathrm{t}}}{I_{\mathrm{0}}}\right)_\lambda = f_{\mathrm{v}} E(m,\lambda) \left(\frac{6\pi  L}{\lambda}\right)
    \label{beerLambertExtinction}
\end{equation}

The spectral dependence of particle extinction is driven in part by its compositional \cite{bescondSootOpticalProperties2016, yonRevealingSootMaturity2021} and nanostructural makeup (amorphous versus ordered graphitic polycrystalline) \cite{michelsenProbingSootFormation2017, gurentsovEffectSizeStructure2022} as discussed in Section \ref{intro}. Thus, the ratio of two extinction measurements from the visible and NIR gives an effective optical maturity ratio:
\begin{equation}
R
=
\frac{\lambda_1 \ln\!\left(\frac{I_t}{I_0}\right)\big|_{\lambda_1}}
     {\lambda_2 \ln\!\left(\frac{I_t}{I_0}\right)\big|_{\lambda_2}}
=
\frac{f_v\,E(m,\lambda_1)}{f_v\,E(m,\lambda_2)}
\label{R_eq}
\end{equation}


\noindent Eq.~(\ref{R_eq}) shows that $R$ is solely a function of the wavelength-dependent absorption functions $E(m,\lambda)$ and independent of $f_\mathrm{v}$. The ratio of extinction at 633 and 1064 nm is therefore used in this study to help describe the evolution of particle maturity, from an organic and amorphous (young) to carbonized and graphitic (mature) carbon particle.

Notably, a wide spread of $E(m)$ is found across various studies due to the complex dependence of the refractive index on particle composition, nanostructure, and size, which in-turn depend on feedstock, gas temperature, and time-dependent evolution of the mixture  (detailed in Section \ref{intro}). To draw conclusions independent of optical property assumptions, extinction results are first reported as relative volume fraction in the form $f_\mathrm{v}E(m,\lambda)$. To infer quantitative $f_\mathrm{v}$ and make comparisons to model predictions, the average of $f_{\mathrm{v,min}}$ and $f_{\mathrm{v,max}}$ were calculated using literature ranges for $E(m)$: $E(m,633) = [0.17, 0.32]$, and $E(m,1064) = [0.20, 0.35]$ \cite{leeOpticalConstantsSoot1981a, liuReviewRecentLiterature2020} with a reported uncertainty that captures the spread in optical properties of $\pm1/2\,(f_{\mathrm{v,max}} - f_{\mathrm{v,min}})$. Measurement-model comparisons made in this work use $\lambda =  1064\ \mathrm{nm}$ to avoid interference from condensing PAHs and the gas medium near 633 $\mathrm{nm}$ \cite{miglioriniInvestigationOpticalProperties2011, russoOpticalBandGap2020}.

The carbon yield ($\mathrm{CY}$) is used to describe the fraction of carbon converted from gas feedstock into the condensed phase. At an assumed particle density $\rho_p$, $\mathrm{CY}$ can be calculated from $f_\mathrm{v}$ by
\begin{equation}
\mathrm{CY}
= 
f_\mathrm{v}(t)\,\frac{\rho_p}{M_C\,[\mathrm{C}]_0} \ \sigma_\mathrm{exp}(t) 
\end{equation}

\noindent where $M_C$ is 12.011 amu, $[\mathrm{C}]_0$ is the initial molar concentration of the carbon atoms at time zero, and $\sigma(t) = \rho_{g,0}/\rho_g(t)$ is a dimensionless expansion factor introduced to account for gas expansion during the reaction time. A particle density $\rho_p$ of 1800\ $\mathrm{kg/m^3}$ was assumed based on an average of available data for soot and CB with different C/H ratios as reported in \cite{adibOmnisootProcessDesign2026}.

\subsection{Sample collection and imaging analysis}
\label{methods_samples}

CB samples were collected using a modified shock tube endwall containing three equidistant plugs located as shown in Fig.~\ref{fig:fig1} with flush-mounted scanning electron microscopy (SEM) stubs (Ted Pella 16111). Carbonaceous particles were deposited thermophoretically onto the stubs during an experiment, then submerged in a methanol bath where the dispersion was transferred onto copper Formvar TEM grids (Ted Pella 018013) via gentle agitation for subsequent TEM imaging. Imaging was performed on an FEI Tecnai G2 F20 X-TWIN TEM, and an FEI Titan TEM (300 kV) to analyze aggregate morphology, primary particle diameter ($d_{\mathrm{p}}$), and internal nanostructure. 

The analysis of TEM images was performed at distinct magnifications to balance population statistics with resolution and detail, utilizing a strategy similar to \cite{scottRapidAssessmentJet2024}. The aggregate morphology was observed at 4,000 – 27,000$\,\times$. For primary particle diameter measurement, images at mid-magnification (100,000$\,\times$) were first analyzed manually to establish \enquote{ground truth} primary particle sizing. Identifiable primary particles were selected at random and manually enclosed with an ellipse using ImageJ software. The mean of the minimum and maximum Feret diameter was chosen to define the $d_\mathrm{p}$ of $\sim$200 individual primary particles per condition. Second, lower magnification (69,000$\,\times$) images were processed using a pretrained neural network for image segmentation (Section \ref{methods_cellpose}). HRTEM images (490,000$\,\times$) were also analyzed to quantify nanostructural metrics, including graphitic layer spacing ($d_\mathrm{f}$), fringe tortuosity $\tau$, and continuous fringe length ($L_\mathrm{f}$) following a method developed by Botero et al. \cite{boteroHRTEMSoot2016} (Section \ref{methods_nanostructure}).

\subsubsection{Neural network-based primary particle segmentation}
\label{methods_cellpose}

The development of rapid techniques to assess crucial morphological metrics of CB has been a longstanding objective \cite{sipkensCharacterizingSootTEM2021, sipkensOverviewMethodsCharacterize2023} \cite{Hess1969CBMorphology}. Primary particle sizes from TEM have been a particularly challenging target to automate due to a high degree of overlap and inconsistent contrast \cite{daySAGEMachineLearning2025}. To scale the analysis of particle size distributions from hundreds of measurements with manual labeling to thousands of data points in this study, automated segmentation was performed using Cellpose-SAM (v4.0.4)  \cite{stringerCellposeGeneralistAlgorithm2021, pachitariu_CellposeSAMS}, a generalizable deep-learning-based segmentation tool originally designed for cellular analysis and recently integrated with the Segment Anything Model (SAM) \cite{pachitariu_CellposeSAMS}. The preprocessing workflow of CB primary particle segmentation using Cellpose-SAM is described in S.M. Section~\ref{SM:cellPose_workflow}. Cellpose-SAM inferences used the \textit{ctyo3} model with an expected diameter of 18 nm, a flow threshold of 0.7, and a cell probability threshold of 0.1. A low cell probability threshold was used to maximize particle detection sensitivity, permitting false positives that can be subsequently filtered during post-processing based on shape descriptors (e.g., circularity, aspect ratio).

A typical TEM image input, (cropped to remove the scale bar before segmentation), with a single particle contour overlaid for visualization is shown in Fig.~\ref{fig:f_cellposeMethods}a. Segmented contour maps were layered over the input image for qualitative assessment in Fig.~\ref{fig:f_cellposeMethods}b, and six effective diameter metrics were applied to each contour (Fig.~\ref{fig:f_cellposeMethods}c) in a given segmentation to compare the bias of each automated approach with manual measurements. These metrics include (1) Inscribed circle (constrained by the nearest boundary indentation, sensitive to surface irregularity); (2) Equivalent circle (area-preserving bulk size metric, commonly used in other CB studies); (3) Equivalent convex hull (area-preserving metric that fills concavities); (4) and (5) Ellipse axes (major and minor axes of the fitted ellipse capture elongation and compactness); and (6) Average Feret diameter (boundary-aware metric representing the average caliper distance, also used in this study for manual measurements).

\subsubsection{Nanostructure image processing}
\label{methods_nanostructure}

\begin{figure}[h]
    \centering
    \includegraphics[width=1.0\linewidth]{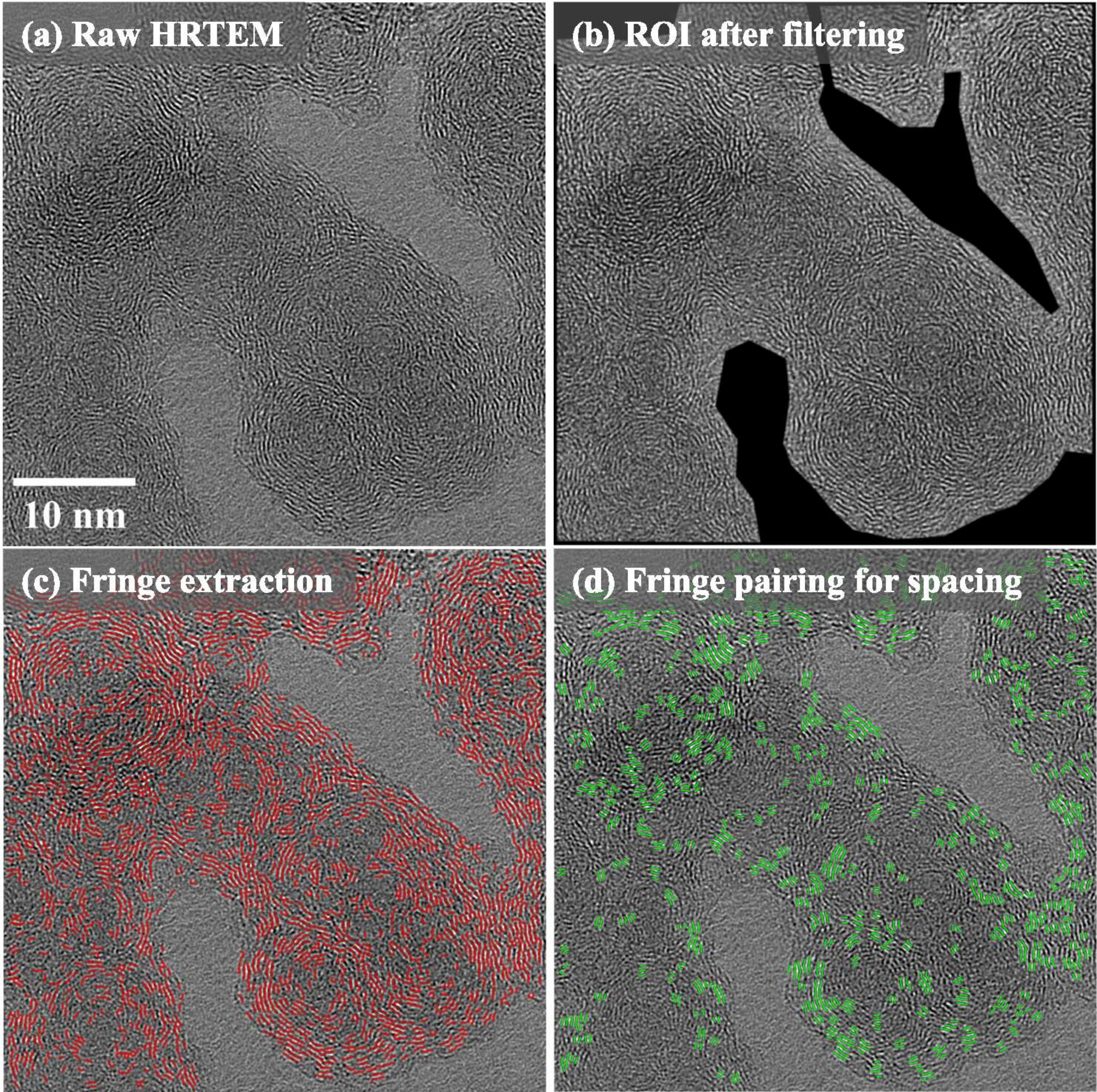}
    \caption{Workflow for quantifying carbon nanostructure fringes from bright-field HRTEM images. (a) Raw micrograph. (b) Selected ROI after preprocessing. (c) Extracted lattice fringes overlaid on the original image (red). (d) Parallel, equal-length fringe pairs within individual stacks (green) used to estimate inter-fringe spacing.}
    \label{fig:fig_fringe_method}
\end{figure}

Lattice-fringe extraction and nanostructure quantification were performed using the open-source MATLAB workflow of Botero et al. \cite{boteroHRTEMSoot2016}, which converts raw bright-field HRTEM micrographs into screened, skeletonized fringe networks from which geometric metrics are computed (Fig.~\ref{fig:fig_fringe_method}). Because fringe statistics can be sensitive to preprocessing choices, a single operator set was selected and held fixed for all conditions to enable consistent cross-condition comparisons; parameter selection and quality-control checks are described in the S.M. In brief, a region of interest (ROI) was defined to exclude non-informative areas (e.g., background, scale bar, and poorly resolved regions), and pixels outside the ROI were set to zero intensity. The ROI was then contrast-enhanced, denoised (Gaussian low-pass), and corrected for non-uniform illumination using a bottom-hat transformation to emphasize dark fringe features prior to binarization (Otsu thresholding). The binary image was skeletonized by iterative thinning to obtain one-pixel-wide centerlines, after which isolated pixels and branched structures were removed using connectivity-based screening so that remaining connected components could be treated as individual fringes \cite{boteroHRTEMSoot2016,yehliuDevelopmentHRTEMImage2011}.

Structural metrics were computed from the screened fringe skeletons. $L_\mathrm{f}$ was obtained by summing pixel-to-pixel distances along each skeleton (including diagonal steps) and converting to physical units using image calibration; $\tau$ was defined as the ratio of fringe arc length to the end-to-end Euclidean distance between fringe endpoints, and $d_\mathrm{f}$ was computed by identifying locally parallel, equal-length fringe pairs within individual stacks (see Fig.~\ref{fig:fig_fringe_method}d) and averaging the minimum separation between paired segments. Physically meaningful lower bounds were imposed as acceptance criteria: $L_\mathrm{f} \ge 0.483$~nm (the length of naphthalene molecules), $\tau \ge 1$ (the tortuosity of straight fringes), and $d_\mathrm{f} \ge 0.3354$~nm (the interlayer spacing of graphite) to target physically meaningful fringe measurements.
 
\section{Simulation}
\label{sim}

Gas-phase chemistry was simulated in Cantera \cite{cantera2023_330} using a zero-dimensional ideal-gas reactor initialized at the post-reflected-shock conditions $(T_5, P_5)$. The experimentally measured pressure (filtered to remove high-frequency noise) was imposed as a time-dependent constraint. Five detailed kinetic models were evaluated against time-resolved measurements and product yields: FFCM-2 \cite{zhangFoundationalFuelChemistry2023}, ABF \cite{appelKineticModelingSoot2000}, CALTECH \cite{blanquartChemicalMechanismHigh2009}, KAUST  \cite{wangPAHGrowthMechanism2013}, and CRECK \cite{pejpichestakulExaminationSootModel2019}. All mechanisms except FFCM-2 include PAH chemistry required to predict CB growth.

Solid particle evolution from CH$_4$ pyrolysis under shock-tube conditions was simulated using Omnisoot, developed in the Energy Particle Technology Laboratory (EPTL) at Carleton University, which couples gas-phase kinetics to soot inception and surface growth submodels and a sectional particle dynamics framework; implementation details are provided in \cite{adibOmnisootProcessDesign2026}. Based on strong agreement with measured CH$_4$-to-C$_2$H$_2$ conversion (Section~\ref{sec:gas_flux}), temperature and gas chemistry were described by CRECK. To avoid double-counting condensed-phase mass, \textit{BIN} species were removed from the CRECK mechanism prior to coupling with Omnisoot. The PAH precursors used to model particle inception include naphthalene (A2, C$_{10}$H$_8$), acenaphthylene (A2R5, C$_{12}$H$_8$; a cyclopenta-fused PAH containing a five-membered ring), phenanthrene (A3, C$_{14}$H$_{10}$), pyrene and acephenanthrylene (A4 and A3R5, both C$_{16}$H$_{10}$ in CRECK), and cyclopentapyrene (A4R5 \ce{C18H10}), spanning two to four aromatic rings. Particle inception was modeled with a temperature-dependent modified E-bridge formation scheme based on Frenklach and Mebel \cite{frenklachMechanismSootNucleation2020}, with implementation details in \cite{adibOmnisootProcessDesign2026}. Surface growth was described by HACA and PAH adsorption, and particle carbonization rates were described by an Arrhenius form in relation to their H/C ratio, detailed in \cite{kholghyCoreShellInternal2016}.

\section{Results and discussion}
\label{results}

\subsection{Measured time-histories}
\label{absorbances}

\begin{figure}[h]
    \centering
    \includegraphics[width=1.0\linewidth]{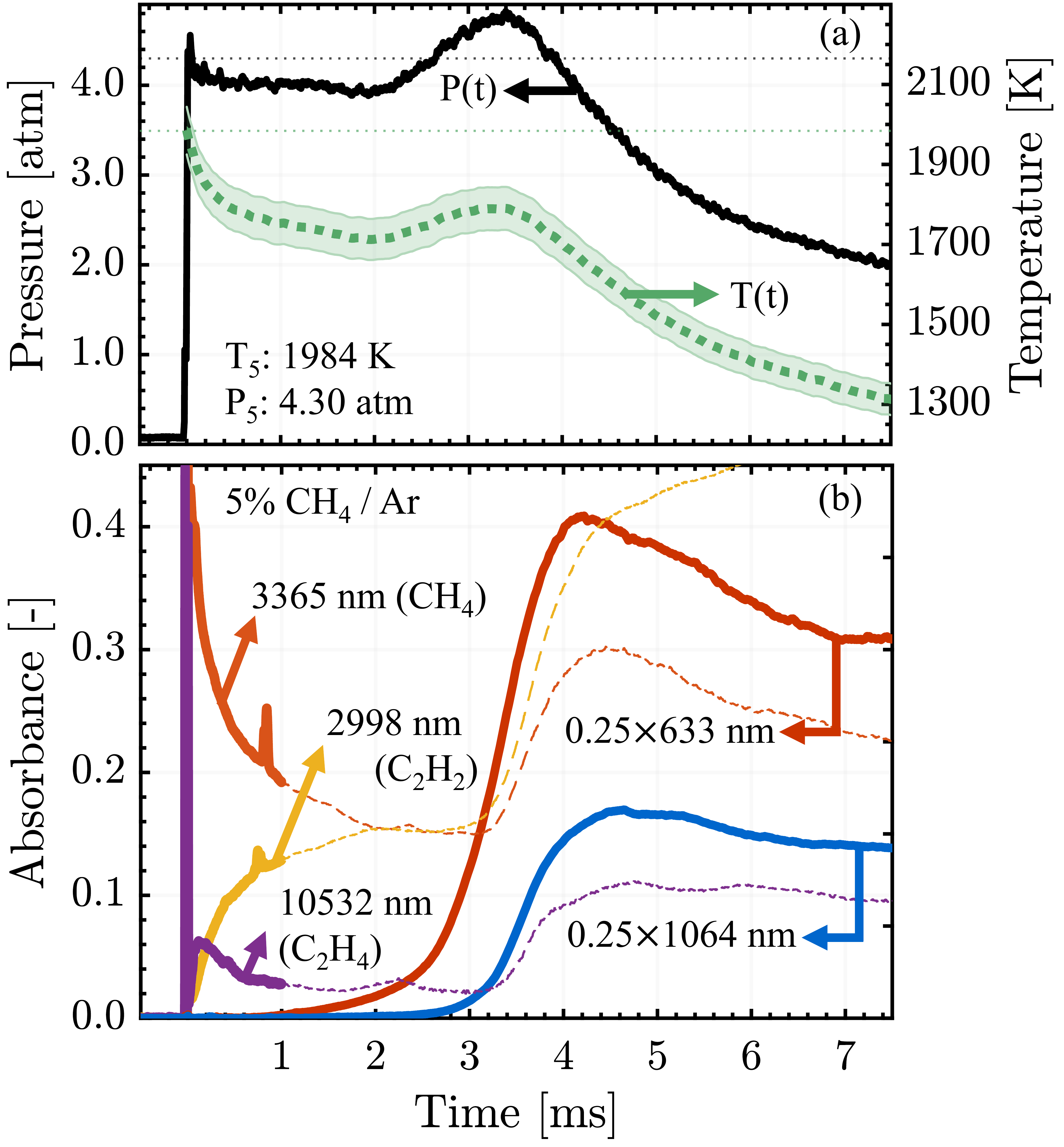}
    \caption{$P(t)$, $T(t)$, gas chemistry and carbon particle formation in CH$_4$ pyrolysis at 1984\,K and 4.3\,atm: (a) $P(t)$ (black) is relatively flat for the first 2 ms, a reflection off of the contact surface drives a rise between 2.5 and 3.5 ms before dropping in the expansion fan. Modeled $T(t)$ (green markers) drops due to endothermic initiation chemistry then follows $P(t)$ isentropically; (b) the reflected shock arrives at time-zero. Measurements of CH$_4$ conversion into C$_2$H$_4$ and C$_2$H$_2$ (thick lines) are made for the first 1 ms (free of particle interference) and extinction at 633 nm (red) and 1064 nm (blue) tracks particle $f_\mathrm{v}$.}
    \label{fig:f_abs_ext}
\end{figure}

\begin{figure*}[h]
    \centering
    \includegraphics[width=1.0\textwidth]{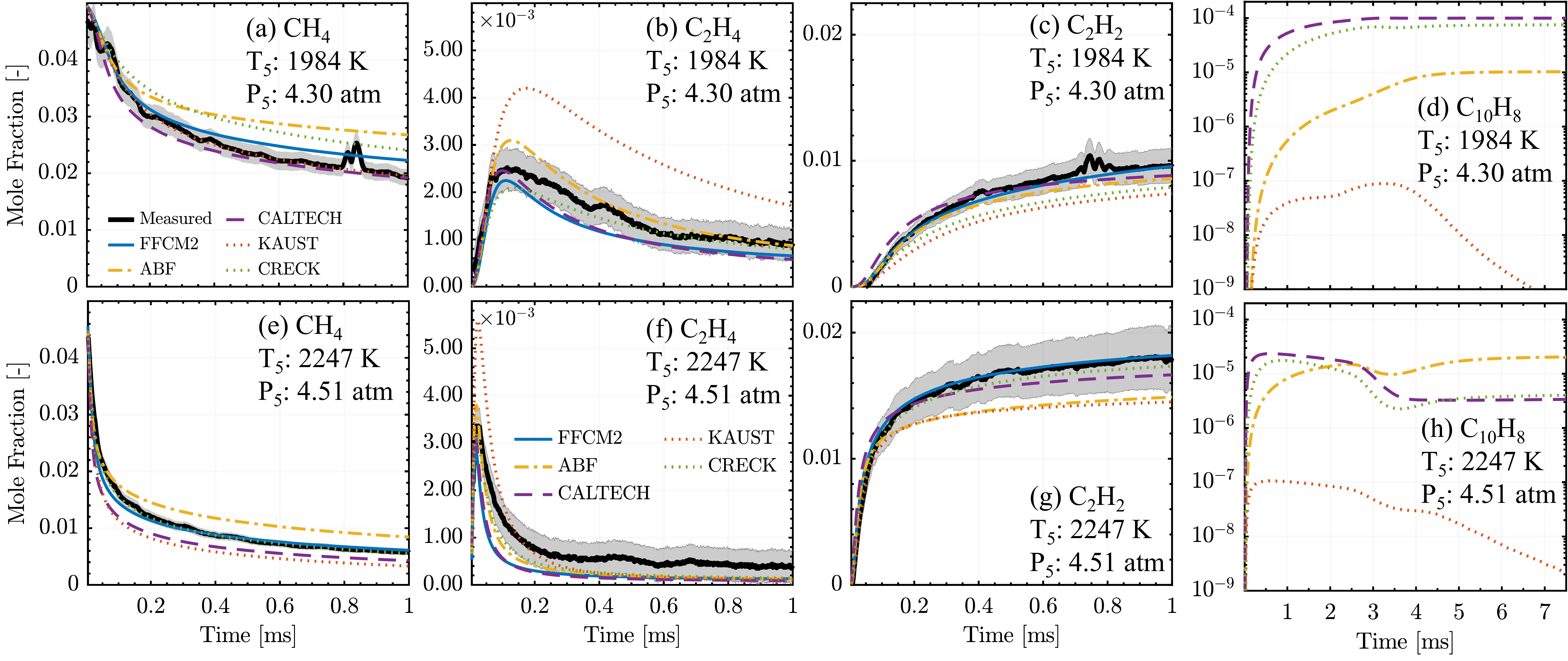}
    \caption{Measured time-histories of CH$_4$ (a, e), C$_2$H$_4$ (b, f), and C$_2$H$_2$ (c, g) compared to kinetic model predictions at $T_5$ = 1984 K (a-d) and 2247 K (e-h). The shaded region around each measurement indicates experimental uncertainty.  Kinetic models FFCM-2, CRECK, and CALTECH show good agreement with the primary gas products measured. Model predictions of PAH concentrations involved in particle inception and surface growth are subject to order-of-magnitude variations as demonstrated by naphthalene (C$_{10}$H$_8$) in d, h.}
    \label{fig:fig_gas_speciation}
\end{figure*}

\begin{figure}[h]
    \centering
    \includegraphics[width=0.85\linewidth]{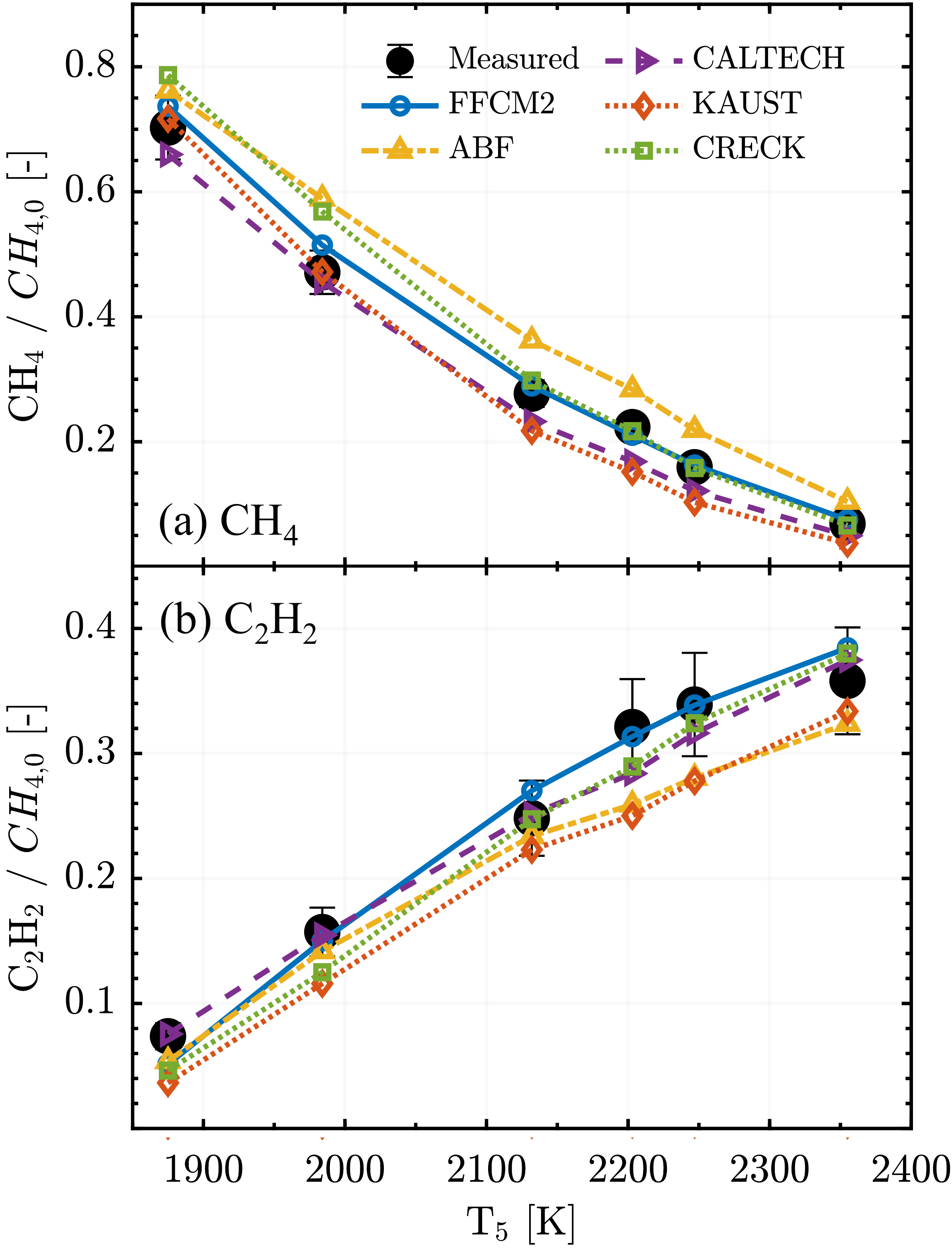}
    \caption{Normalized yields ($X_\mathrm{i}(t)/X_{\mathrm{CH_4},0}$) measured at $t$ = 0.5 ms for (a) CH$_4$ conversion into (b) C$_2$H$_2$, compared with kinetic simulations for $T_5$ from 1875\,K to 2355\,K. C$_2$H$_2$ rises monotonically with $T_5$. Most kinetic models predict small hydrocarbon yield well. Errors bars indicate experimental uncertainty, dashed lines are added for clarity in observed trends.}
    \label{fig:fig_gas_yields}
\end{figure}

\begin{figure}[h]
    \centering
    \includegraphics[width=1.0\linewidth]{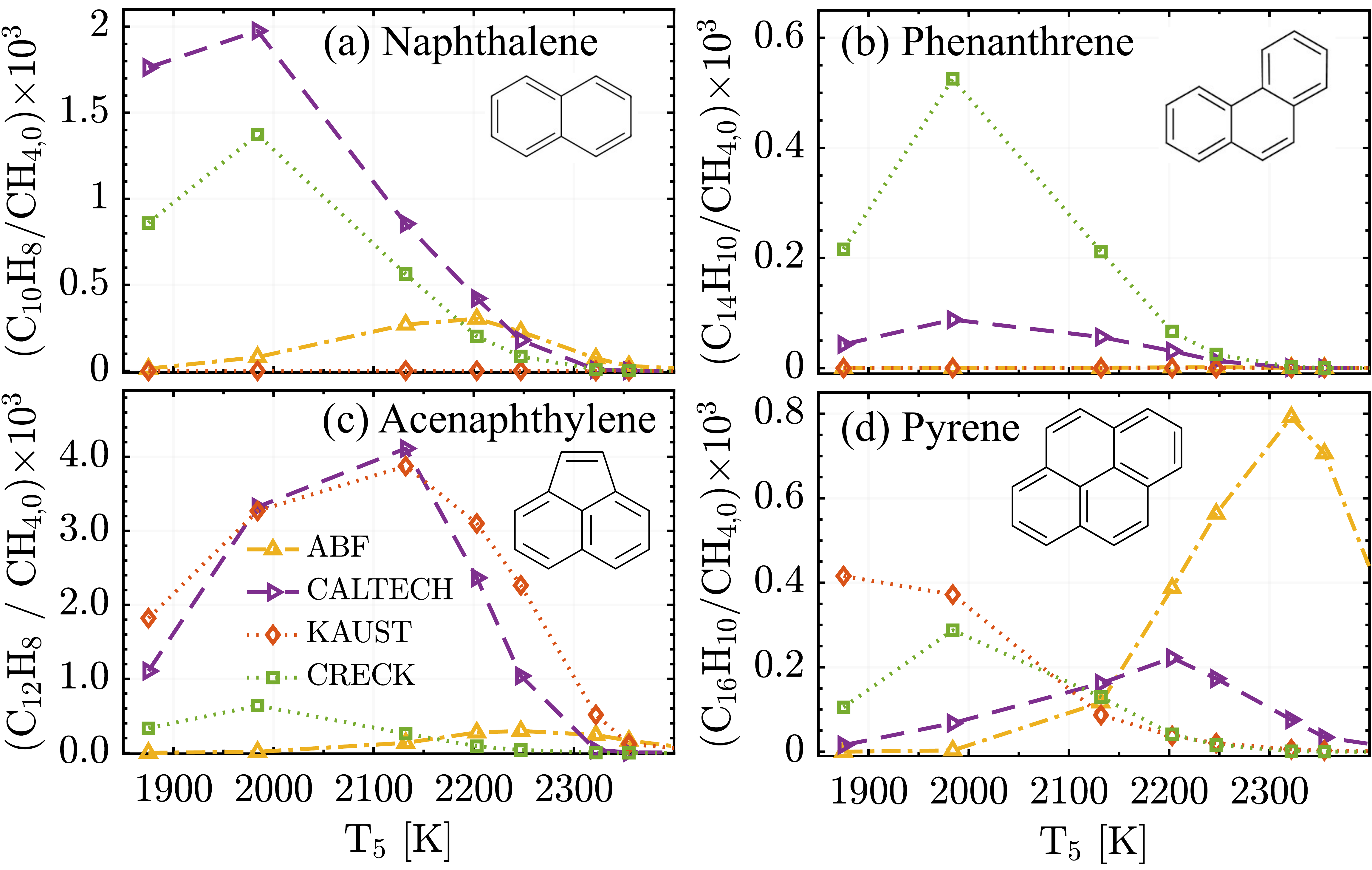}
    \caption{Normalized PAH yields ($X_\mathrm{PAH}(t)/X_{\mathrm{CH_4},0}$) vs $T_5$ at $t = 3.0$ ms predicted by CRECK, CALTECH, KAUST, and ABF: (a) naphthalene (C$_{10}$H$_8$), (b) phenanthrene (C$_{14}$H$_{10}$), (c) acenaphthylene (C$_{12}$H$_8$), and (d) pyrene (C$_{16}$H$_{10}$). The predicted PAH precursor pool is strongly mechanism-dependent in both magnitude and temperature, in contrast to the monotonic trends observed for the major products.}
    \label{fig:fig_pah_yields}
\end{figure}

Figure~\ref{fig:f_abs_ext}a shows the measured pressure, $P(t)$, and the corresponding temperature simulations, $T(t)$, during the pyrolysis of 5\% CH$_4$/Ar behind a reflected shock at $T_5 = 1984$~K and $P_5 = 4.30$~atm. $T(t)$ is simulated with the CRECK model, (which agrees very well with $T(t)$ predictions from FFCM-2). A 3\% uncertainty conservatively captures the spread in $T(t)$ from all kinetic models. Simulations follow the prescribed experimental $P(t)$ to capture the influence of compression and expansion on the experimental temperature. One representative time-history is shown in Fig. \ref{fig:f_abs_ext} for brevity. Results from $T_5=$ 1850 to 2450~K ($P_5 = 4.5 \pm 0.4$~atm), are discussed below. The canonical test time associated with a steady pressure trace extends to around 2~ms, followed by a pressure rise near 3~ms due to shock reflection off of the driver-gas contact surface and subsequent cooling caused by the arrival of the expansion fan. In the first 1~ms, $T(t)$ drops nearly 200\,K due to the highly endothermic initiation step (R1) and dehydrogenation steps leading to C$_2$H$_2$ formation. The subsequent heat release from later-stage chemistry and particle growth is negligible (in 95\% Ar) compared to pressure variations; consequently, the temperature evolves approximately isentropically with pressure.

Simultaneously, absorbance time-histories capturing the conversion of CH$_4$ into C$_2$H$_4$, C$_2$H$_2$, and condensed-phase particulate are measured by laser absorption and light extinction (Fig.~\ref{fig:f_abs_ext}b). The induction period between fuel pyrolysis at time zero, small-hydrocarbon chemistry, PAH growth, and particle formation is reflected by the 1--2~ms delay before measurable light extinction at 633 and 1064~nm. Absorbance data are therefore conservatively restricted to the first 1~ms to infer gas-phase species mole fractions and make model comparisons free from particle interference (Section~\ref{sec:gas_flux}), while extinction data are processed to infer time-resolved optical maturity, $f_\mathrm{V}$, and global properties such as induction time, ($\tau_{\mathrm{ind}})$, and $\mathrm{CY}$ (Section~\ref{sec:time_resolved_synthesis}). Measurements into the expansion fan enable the evaluation of model predictions of CB growth and morphology that incorporate time-variation in reactor pressure.

\subsection{Gas phase carbon flux}
\label{sec:gas_flux}

Figures~\ref{fig:fig_gas_speciation}a-h show measured mole fractions and kinetic model comparisons for carbon flux from CH$_4$, C$_2$H$_4$, and C$_2$H$_2$ into larger PAHs (e.g. C$_{10}$H$_8$) at two representative temperatures (1984 K, and 2247 K). At $T_5=$1984 K, CH$_4$ conversion reaches about 60\% by 1 ms (Fig.~\ref{fig:fig_gas_speciation}a), while at 2247 K, 90\% of fuel is consumed by 1 ms (Fig.~\ref{fig:fig_gas_speciation}e).  Consistent with known CH$_4$ pyrolysis pathways, C$_2$H$_4$ forms rapidly and is subsequently depleted, with its peak narrowing substantially as temperature increases (Figs.~\ref{fig:fig_gas_speciation}b,f). C$_2$H$_2$ forms quickly and remains the most abundant unsaturated hydrocarbon product over the measurement time (Figs.~\ref{fig:fig_gas_speciation}c,g). Measurement uncertainties reflect the propagation of errors in temperature, pressure, and absorption coefficients. Uncertainty is the largest for C$_2$H$_2$ due to its spectral sensitivity to interfering CH$_4$ (mitigated by measuring and subtracting interference as discussed in Section~\ref{methods_LAS}). There is a narrow spread in model predictions of small hydrocarbon chemistry.  

Figures~\ref{fig:fig_gas_speciation}d,h depict naphthalene (C$_{10}$H$_8$) time-histories predicted by the four models containing aromatic pathways. Predictions of PAH growth into later times show orders of magnitude disagreement across models, (note the logarithmic scale for Figs.~\ref{fig:fig_gas_speciation}d,h). KAUST predicts systematically low C$_{10}$H$_8$, ABF is the most temperature sensitive and accumulates C$_{10}$H$_8$ at higher $T_5$, while in both CALTECH and CRECK, the high $T_5$ behavior indicates that C$_{10}$H$_8$ acts as transient intermediate in aromatic growth: at early times it is formed, at intermediate times it is converted along aromatic channels in the model, and at later times the expansion fan and cooling allow partial re-accumulation. Coupling this detailed chemistry to Omnisoot adds particle inception and surface growth as additional sinks for the PAHs designated as precursors.

Temperature-dependent 0.5 $\mathrm{ms}$ yields normalized by fuel loading (CH$_{4,0}$) reinforce these trends between 1875\,K and 2355\,K (Fig.~\ref{fig:fig_gas_yields}). Increasing $T_5$ monotonically increases fuel conversion, favoring higher C$_2$H$_2$ yields, and the detailed models show comparatively small spread for these major-species. CRECK and CALTECH provide the closest overall agreement with measurements across all $T_5$, whereas ABF and KAUST underpredict C$_2$H$_2$ yield. Overall, the small-hydrocarbon pool is well-constrained by the measurements and models, and the largest uncertainty in particle-relevant gas chemistry enters primarily through the pathways governing aromatic growth.

In Fig.~\ref{fig:fig_pah_yields}, yields of four of the PAHs designated as particle precursors exhibit a loose bell-shaped temperature dependence with varying peak temperatures. CALTECH and CRECK produce the largest naphthalene and phenanthrene yields at low $T_5$ but decrease with increasing $T_5$ (Figs.~\ref{fig:fig_pah_yields}a,b), whereas KAUST produces negligible naphthalene and phenanthrene while generating substantially higher acenaphthylene (Fig.~\ref{fig:fig_pah_yields}c). In contrast, ABF predicts relatively low acenaphthylene, but the highest pyrene yields at the highest $T_5$ (Fig.~\ref{fig:fig_pah_yields}d). The disagreement among models poses a target for more measurements and model refinement and highlights the challenge that PAH predictions pose to models of CB formation and growth across a range of temperatures.

Because C$_2$H$_2$ is a primary reactant in PAH and particle surface growth pathways, agreement in CH$_4$ to C$_2$H$_2$ conversion provides an important constraint for modeling subsequent particle synthesis and motivates the choice of a gas-phase chemical mechanism for such modeling. FFCM-2 reproduces the major pyrolysis products well but does not contain aromatic pathways needed to represent particle inception and growth. Among the large mechanisms considered, CRECK agrees well with the measured conversion of CH$_4$ to C$_2$H$_2$ and was chosen as the base chemistry for detailed particle modeling.

\subsection{Time resolved particle synthesis}
\label{sec:time_resolved_synthesis}

\subsubsection{Multiwavelength light extinction}
\label{sec:multiwavelength_LEx}

\begin{figure}[h]
    \centering
    \includegraphics[width=0.90\linewidth]{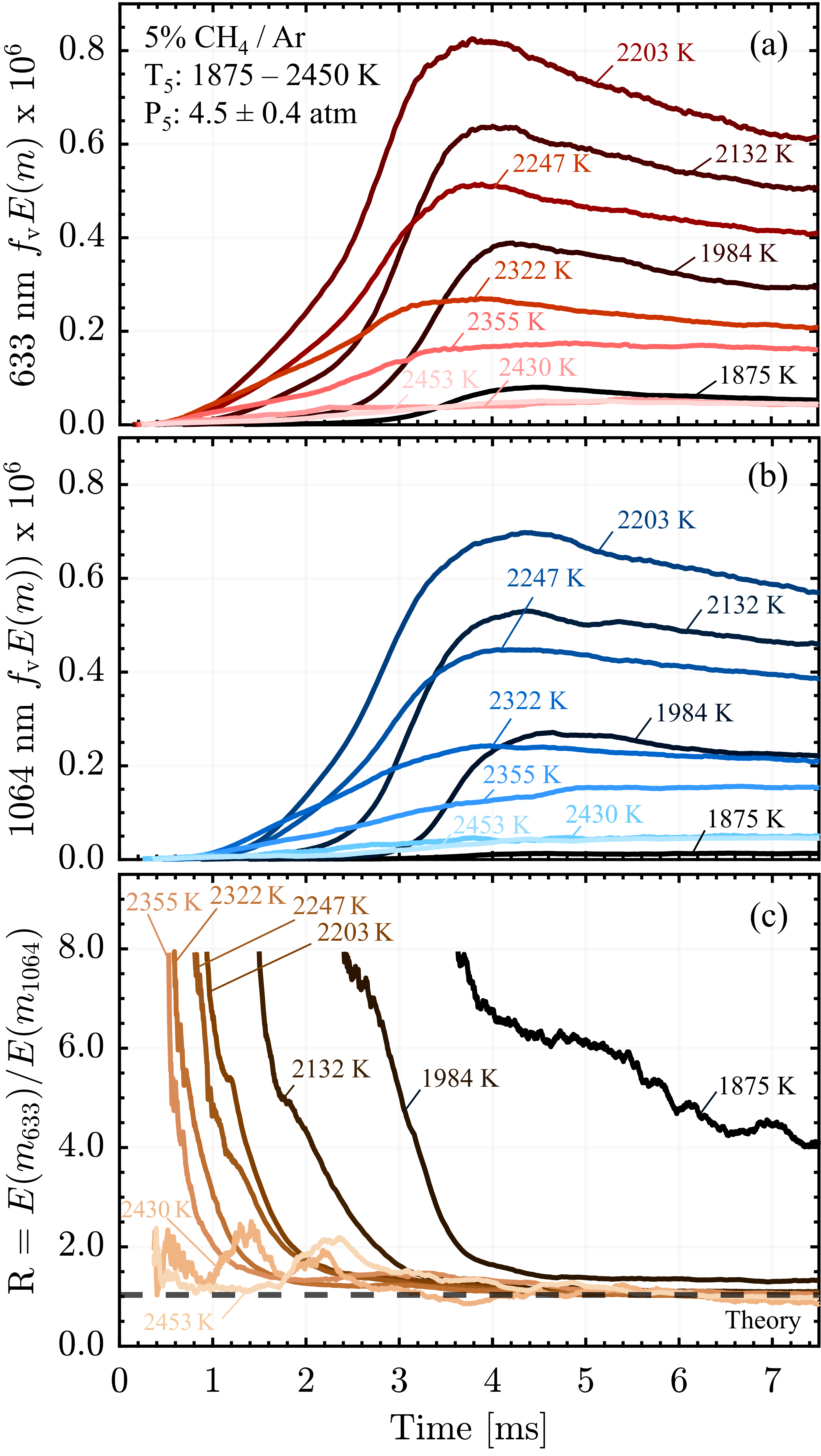}
    \caption{Light extinction measurements at 633 nm (a) and 1064 nm (b) for $T_5$ from 1875\,K to 2453\,K. The corresponding evolution of the optical maturity ratio $R = E(m,633_{\mathrm{nm}})/E(m,1064_{\mathrm{nm}})$ in (c) is compared to the constant dashed line for mature particles from LII in flames \cite{bejaouiExperimentalStudyEm2014}}
    \label{fig:fig_LEx}
\end{figure}

Time-resolved light extinction at multiple wavelengths was measured to quantify the onset and evolution of condensed-phase particles during CH$_4$ pyrolysis. Extinction was measured simultaneously at 633 and 1064~nm to provide complementary sensitivity to young amorphous vs. more carbonized and graphitic material. The resulting multiwavelength dataset enables extraction of (1) extinction-ratios $R(t)$ that are indicative of changing particle optical properties, (2) time-resolved $f_{\mathrm{v}}(t)$, and (3) integrated metrics including $\tau_{\mathrm{ind}}$, and CY, which are used to evaluate detailed numerical simulations in Omnisoot.

Figures~\ref{fig:fig_LEx}a,b report relative volume fractions, $f_{\mathrm{v}}E(m)$, to provide an interpretation that is independent of absorption function assumptions and to highlight the dominant growth regimes for carbon particles. These regimes include: (1) an temperature-dependent induction period between CH$_4$ thermal decomposition and particle inception; (2) a rapid increase in carbon flux into particles driven by surface growth; and (3) subsequent coagulation and dehydrogenation toward mature particulate. This interpretation is supported by detailed model calculations in a companion publication currently in preparation by Adib et al. \cite{adibClark_ppr2}. The noise floor for all extinction measurements was below $f_{\mathrm{v}}E(m)=2\times 10^{-9}$ at both 633 and 1064~nm. The strong temperature dependence of formation rates, yields, and induction times (defined in various ways in the literature) are discussed in the following sections.

Extinction signals in Fig. \ref{fig:fig_LEx} decay at late times because gas expansion in the shock tube increases the denominator of $f_\mathrm{v}=V_{p}/V_{gas}$. In most shock tube particle synthesis studies, reported data are thus restricted to a few milliseconds for the purpose of assuming constant pressure behind the reflected shock wave \cite{kcSimultaneousMeasurementsAcetylene2017, ereminFormationCarbonNanoparticles2012, agafonovSootFormationPyrolysis2015}. This approach is appropriate when (a) experimental care is taken to ensure the pressure is indeed constant over the analysis window, and (b) particle inception and growth are sufficiently fast to occur within this time (e.g., in aromatic mixtures). In contrast, saturated hydrocarbons are slower to form particles \cite{ereminFormationCarbonNanoparticles2012, kellererMeasurementsGrowthCoagulation2000} and such limited-time data can risk neglecting essential growth information. Light extinction was therefore analyzed beyond the canonical test time and into the expansion fan to build a more complete picture of nanoparticle formation from CH$_4$ pyrolysis. This strategy also supports a broader aim of improving and evaluating predictive capabilities for CB synthesis under dynamic reactor conditions.

Taking the ratio of extinction signals at different wavelengths removes dependence on $f_{\mathrm{v}}$ and gives the time-dependent optical maturity ratio $R(t)$ (Fig.~\ref{fig:fig_LEx}c). $R(t)$ is initially elevated and then decays toward a plateau near unity, with the decay occurring more rapidly at higher $T_5$. Consequently, incomplete decay is observed at $T_5=1875$~K due to insufficient energy to generate particles of spectrally independent refractive indices. The decay in $R(t)$ is consistent with a transition from a more organic, amorphous, and less carbonized material toward increasingly carbonized and graphitic (mature) particles with reduced spectral contrast between 633 and 1064~nm \cite{johanssonEvolutionMaturityLevels2017, cepedaPonMaturity2025}. This behavior indicates that particle maturity evolves measurably over millisecond timescales and is highly sensitive to temperature (temperature time-histories can be found in the S.M.). The systematic time and temperature dependence of $R(t)$ further indicate that an assumption of constant $E(m,\lambda)$ at early times and lower temperatures in CH$_4$ pyrolysis should be treated with substantial uncertainty. Increased noise in $R(t)$ above 2400\,K is caused by division of two signals with low signal-to-noise ratios (SNR) due limited particle formation. Finally, the relative timing of the plateau in $R(t)$ compared to the continued growth of $f_{\mathrm{v}}E(m,\lambda)$ at a given temperature suggests either (a) mechanisms for mass growth continue beyond the time at which structural and compositional changes governing optical maturity have stabilized, or (b) that particles with mature optical properties dominate the path-integrated extinction signal even while a population of young particles persists. The extent of particle carbonization at each condition was investigated with TEM to explore these maturity hypotheses in Section \ref{sec:carbon_particle_nanostructure_analysis}.

\subsubsection{Particle volume fractions and model comparisons}
\label{sec:solid_particle_volume_fractions}

\begin{figure}[h]
    \centering
    \includegraphics[width=1.0\linewidth]{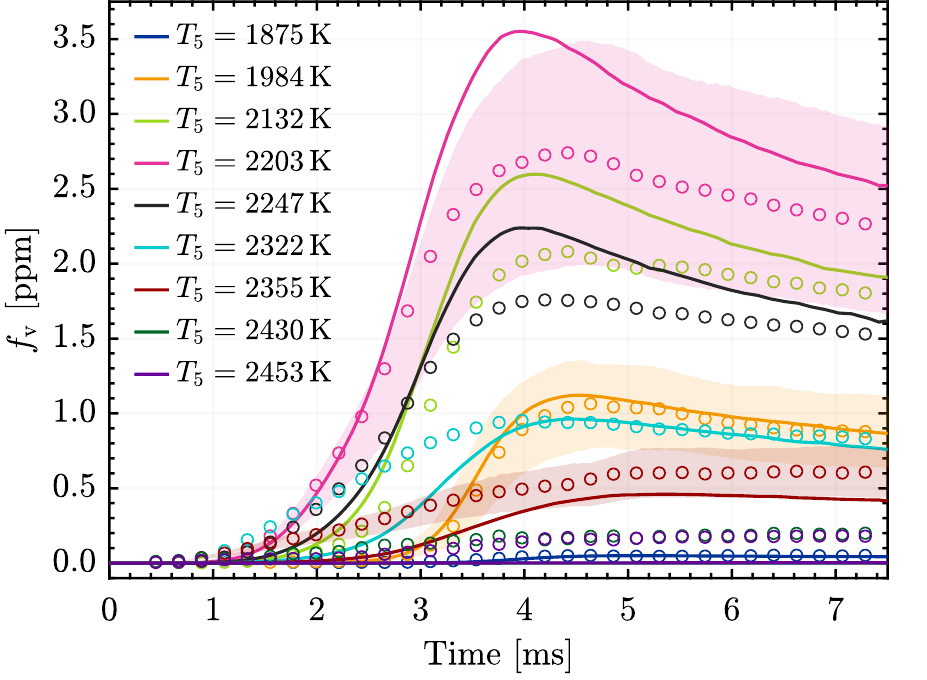}
    \caption{Measurements (markers) at 1064  nm and Omnisoot predictions (lines). For legibility shaded error bars that represent the uncertainty in $E(m)$ of evolving carbon particulate are shown only for $T_5$ = 1984\,K, 2203\,K, and 2355\,K. The best agreement is seen for 1875\,K (blue) and 1984\,K (orange), while $f_\mathrm{v}$ is overestimated at mid-$T_5$ and underestimated at higher temperatures.}
    \label{fig:fig_fv}
\end{figure}

Extinction at 1064~nm was converted to condensed phase volume fraction, $f_{\mathrm{v}}(t)$, using Eq.~(\ref{beerLambertExtinction}) and a wide bound of $E(m,1064)$ from the literature to represent uncertainty introduced by the constant $E(m)$ assumption (Section~\ref{methods_LEx}). Figure~\ref{fig:fig_fv} compares the measured $f_{\mathrm{v}}(t)$ time-histories to Omnisoot predictions across the temperature range studied. It should be noted mature particles (which dominate at later times) possess higher $E(m)$ which is reflected by the lower error bound in Fig.~\ref{fig:fig_fv}. 

Particle $f_{\mathrm{v}}$ formation was observed to increase quickly with temperature from 1875~K to a peak at 2203~K. Beyond 2203~K, formation rates and yield decrease. While there are various potential descriptions of why this occurs \cite{ereminFormationCarbonNanoparticles2012}, this temperature trend is largely responsible for the well-known ``soot bell curve'' and matches the temperature trends of many of the likely PAH precursors (Fig.~\ref{fig:fig_pah_yields}). This decrease in formation beyond 2200~K will also be discussed in conjunction with temperature-dependent induction times (Section~\ref{sec:particle_synthesis_induction_times}).

Omnisoot predictions agree in relative formation rates and magnitudes with measured particle growth, although the best agreement is found for low temperature ($T_5 = 1875$ \& $1984$~K). Because this agreement is achieved through combined optimization of particle inception and surface growth rates to minimize predictive error, it is important to constrain and evaluate the model with targets beyond a simple yield or short time history \cite{adibOmnisootProcessDesign2026}. Constraining the model with the measured $P(t)$ enables an accurate prediction of the decaying $f_{\mathrm{v}}$ due to gas expansion. Model--measurement deviations show three consistent behaviors. First, onset timing errors become more pronounced at higher $T_5$ (quantified more directly by induction times in Fig.~\ref{fig:fig_ind}). Second, Omnisoot tends to overestimate the peak $f_{\mathrm{v}}$ from 2100-2250 K. Third, Omnisoot underestimates $f_{\mathrm{v}}$ at the highest $T_5$ conditions, suggesting that the temperature dependence of net carbon flux into the condensed phase (via inception and surface growth) is not captured adequately at high temperatures. These results demonstrate that the model that captures temperature-dependent carbon particle synthesis from \ce{CH4} behind reflected shocks, but needs updated descriptions of inception and growth pathways that scale to higher temperatures.

\subsubsection{Induction times of particle formation}
\label{sec:particle_synthesis_induction_times}

\begin{figure}[h]
    \centering
    \includegraphics[width=1.0\linewidth]{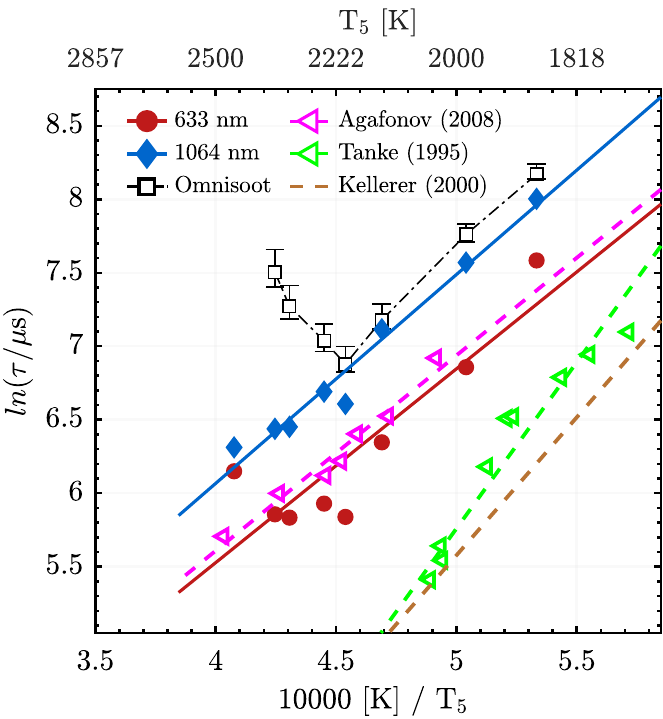}
    \caption{Induction times, $\tau_{\mathrm{ind}}$, extracted from $f_\mathrm{v}E(m,\lambda)$ extinction signals using a noise-floor crossing criterion ($\num{2e-9}$) for 633~nm (red circles) and 1064~nm (blue diamonds) and compared to Omnisoot predictions (squares) where $E(m) = 0.2$ and error bars represent the spread in predictions with $ 0.1 \leq E(m) \leq 0.3$. Additional studies of CH$_4$ in shock tubes include Agafonov (2008) 5\% CH$_4$/Ar at $\sim$5~atm (magenta triangles) \cite{agafonovSootFormationPyrolysis2008}, Tanke (1995) 1--2\% CH$_4$ at 25--54~atm (green triangles) \cite{tankeRussbildungKohlenwasserstoffpyrolyseHinter1994}, and Kellerer (2000) CH$_4$ oxidation modeling (brown dashed) \cite{kellererMeasurementsGrowthCoagulation2000}. Lines denote least-squares fits to the data and slopes provide an effective activation energy, $E_{\mathrm{ind}}$.}
    \label{fig:fig_ind}
\end{figure}

Induction times, $\tau_{\mathrm{ind}}$, provide a compact metric for comparing the temperature dependence of particle inception between experiments and models. The value of $\tau_{\mathrm{ind}}$ depends partly on the experimental technique used \cite{knorreSootFormationPyrolysis1996}. In this study, we define $\tau_{\mathrm{ind}}$ as the time at which the signal crosses a noise floor of $f_{\mathrm{v}}E(m)=2\times 10^{-9}$ (set by beam steering and detector noise) to best represent the delay between fuel decomposition and the first detected particle formation. This threshold corresponds to $13\pm7$ ppb, depending on the assumed $E(m)$. The same approach has recently been applied in studies of C$_2$H$_4$ and C$_6$H$_6$ pyrolysis \cite{kcSimultaneousMeasurementsAcetylene2017} to minimize sensitivity to the slope of the growth curve which can be strongly influenced by pressure gradients. A second common definition in the literature is the intersection of the tangent to the $f_{\mathrm{v}}$ signal inflection point with the abscissa. A comparison of induction times extracted using the noise-floor and inflection-tangent approaches for the present data is provided in Section~\ref{SM:indTimes} of the S.M.

Noise floor induction-times from Omnisoot are largely insensitive to uncertainty in $E(m)$. Here, $\tau_{\mathrm{ind}}$ is defined in $f_{\mathrm{v}}E(m)$ space, and Omnisoot outputs were mapped to the same $f_{\mathrm{v}}E(m)$ representation using $E(m) = 0.20$ to reflect incipient particles \cite{kelesidisDeterminationVolumeFraction2021}. Uncertainty is estimated by conservatively varying $E(m)$ from $0.10$ to $0.30$, yielding an inferred spread of $\Delta\tau \approx 300~\upmu\mathrm{s}$, corresponding to a $1-3\%$ uncertainty in $\ln(\tau/\upmu\mathrm{s})$ space. The temperature dependence of measured induction times (Fig.~\ref{fig:fig_ind}) is well fit by an Arrhenius expression:

\begin{equation}
    \tau_{\mathrm{ind}} = A_{ind}\,\exp\!\left(\frac{E_{\mathrm{ind}}}{RT_5}\right)
    \label{eq:arrhenius_tau_ind}
\end{equation}

where $E_{\mathrm{ind}}$ is the apparent activation energy associated with the net inception pathways controlling the onset of detectable condensed-phase formation.

Figure~\ref{fig:fig_ind} summarizes induction times on an Arrhenius plot, comparing 633~nm and 1064~nm signals to past studies and Omnisoot predictions. Two experimental induction-time measurements are reported because they probe different effective sensitivities. The 633~nm results match prior shock-tube CH$_4$ pyrolysis studies at the same wavelength by Agafonov (2008) \cite{agafonovSootFormationPyrolysis2008} at similar dilution and pressure, and are systematically faster than 1064~nm, consistent with 633~nm detecting less-mature material with higher aromatic content. The $E_\mathrm{ind}$ suggested by the slopes of both 633 and 1064 nm agree well with \cite{agafonovSootFormationPyrolysis2008}. Other induction times in the literature are likely faster due to different stoichiometry and pressure: Tanke (1995) \cite{tankeRussbildungKohlenwasserstoffpyrolyseHinter1994} measured pyrolysis of 1--2\% CH$_4$ at 25--54~bar, and Kellerer et al. (2000) \cite{kellererMeasurementsGrowthCoagulation2000} modeled soot formation from CH$_4$ oxidation.

Omnisoot induction times align more closely with the slower 1064~nm induction times than with 633~nm at lower and mid-$T_5$. At higher temperatures (above $\sim$2200~K), the model diverges from the observed induction behavior, predicting slower induction at higher $T_5$. The reported Omnisoot trend is independent of the choice between the noise floor or inflection tangent analysis method. This indicates that inception chemistry, including the buildup of the precursor pool feeding inception and the inception pathways themselves remains a subject of further improvement. Martin et al. (2022) \cite{martinSootInceptionCarbonaceous2022} recently reported that a combination of chemical and physical mechanisms may be required to adequately describe such high-temperature inception. Interestingly, this non-Arrhenius behavior was observed at 633~nm at substantially higher $T_5$ (2453~K). This high-$T_5$ measurement can be interpreted as the result of diminished total formation such that, in the limit of high $T_5$, it can take exceedingly long for the signal to cross the noise floor. In fact, Omnisoot predictions never exceed the noise floor at this $T_5$. The rapid maturation driven by extremely high $T_5$ (Fig.~\ref{fig:fig_LEx}c) could also contribute to the lack of delay between 633 and 1064 nm, while different methods to extract induction times likely contribute to the deviation from \cite{agafonovSootFormationPyrolysis2008} at 2453 K. Together, these comparisons show that induction time can help assess PAH-driven inception assumptions and their modeled temperature dependence during the thermal pyrolysis of CH$_4$.

\subsubsection{Carbon yields}
\label{sec:volume_fraction_yields}

\begin{figure}[h]
    \centering
    \includegraphics[width=1.0\linewidth]{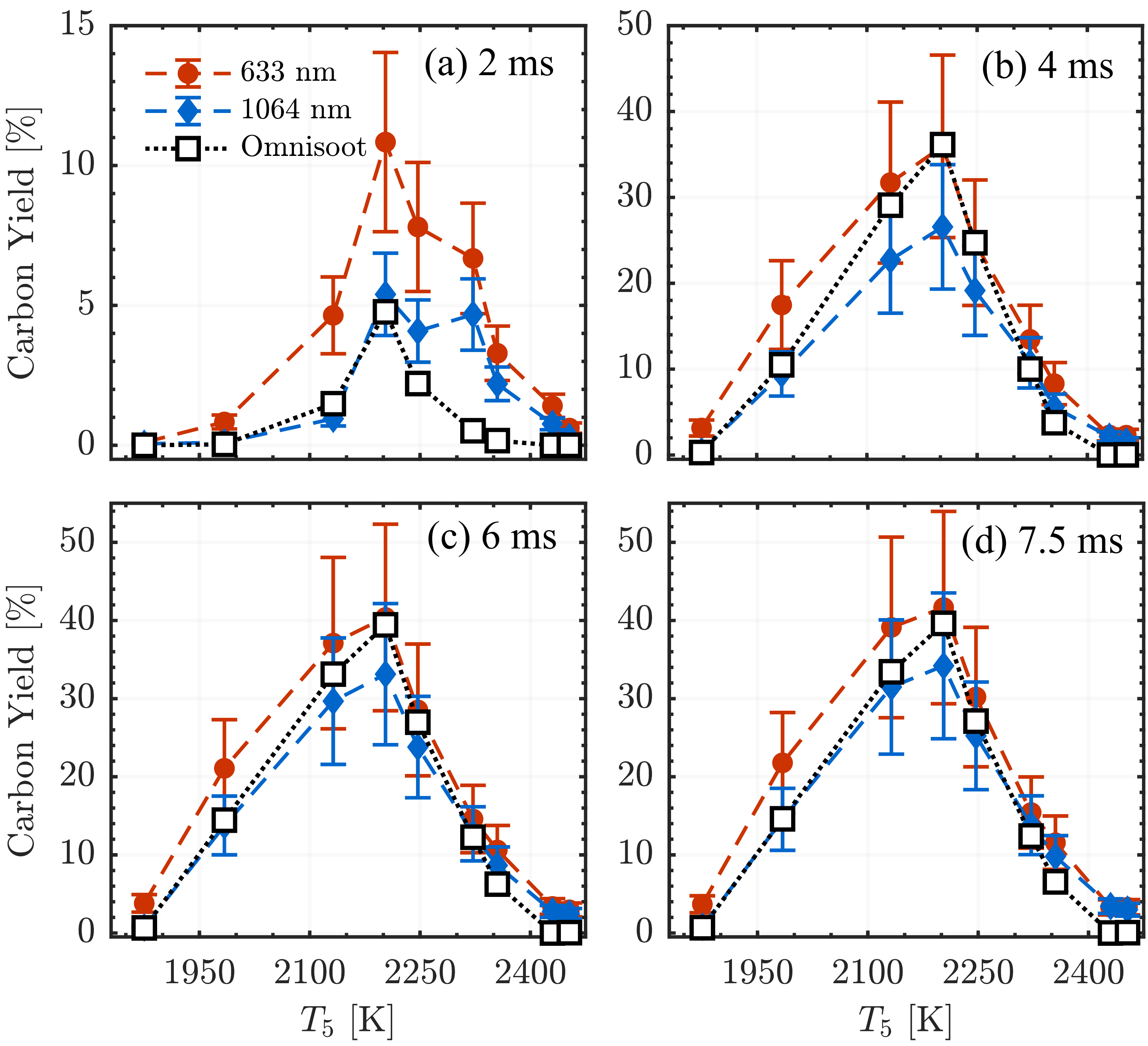}
    \caption{Carbon yield measured at 633 nm (red circles) and 1064 nm (blue diamonds) sampled at (a) 2 ms, (b) 4 ms, (c) 6 ms, and (d) 7.5 ms compared with predictions from Omnisoot (squares). Experimental uncertainty is due to variability in $E(m)$ and a dashed line is added to show trends.}
    \label{fig:fig11_CYields}
\end{figure}

Figure~\ref{fig:fig11_CYields} compares $\mathrm{CY}$ sampled at fixed times (2, 4, 6, and 7.5~ms) across $T_5$ for both wavelengths against Omnisoot predictions. Error bars reflect a spread in $E(m)$ (where a larger $E(m)$ returns a lower CY, and a smaller $E(m)$ returns a higher CY). At all times, results show the well-known bell-shaped temperature dependence characterized by a peak at 2200\,K followed by lower yields at the hottest conditions. This behavior is consistent with the time-history trends in Fig.~\ref{fig:fig_fv} and shows that maximum condensed-phase production from shock-pyrolysis of $5\%$ CH$_4$/Ar reaches 30-40\% and occurs near a mid-$T_5$ window around 2200 K. At early time (2~ms), the model tracks the 1064~nm yields better than 633~nm, in agreement with the induction-time result of a slower $\tau_\mathrm{ind}$. From 4.0-7.5 ms, Ommisoot overpredicts CY at intermediate $T_5$, but aligns well with 1064 nm at lower and higher temperature. Notably, while the predicted $\tau_\mathrm{ind}$ temperature trend is too slow above 2200 K, Omnisoot does capture the temperature dependence of declining yields well. At later times, the partial convergence of 633 nm and 1064 nm yields is in-part caused by the measured population more closely resembling mature particles, while the different spreads in wavelength-dependent optical-property assumptions used for $E(m,\lambda)$ are partly responsible for residual wavelength-to-wavelength differences.

\subsubsection{The influence of shock tube dynamics}
\label{sec:shock_tube_dynamics}

\begin{figure}[h]
    \centering
    \includegraphics[width=1.0\linewidth]{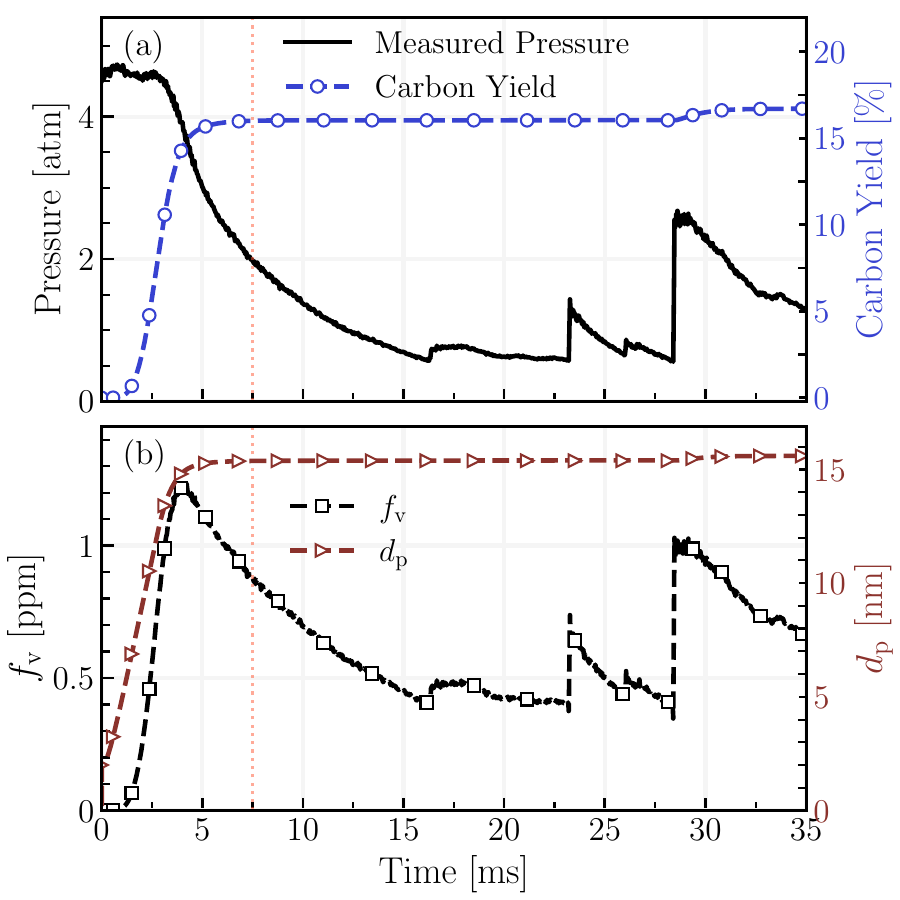}
    \caption{Simulating the influence of subsequent compression and expansion waves that result from shock tube gas dynamics at long times on carbon conversion and morphology: (a) Measured $P(t)$ for 35 ms after a representative shock ($T_5=$ 2238\,K / $P_5=$ 4.5 atm) in FAST (black) and the resulting carbon yield prediction from Omnisoot using a monodisperse particle dynamics model (blue). (b) Simulated $f_\mathrm{v}$ (black) and $d_\mathrm{p}$ (brown) in the same condition. Omnisoot predicts post-7.5 ms evolution of $\Delta \mathrm{CY}$ =0.0069 (4.34\%) and $\Delta d_\mathrm{p}$ = 0.2185 nm (1.42\%) from 7.5 to 35 ms. }
    \label{fig:f_longt_ext}
\end{figure}

\textit{in situ} laser diagnostics constrain particle evolution to $ 7.5\,\mathrm{ms}$, yet post-shock sample collection for electron beam microscopy necessarily integrates over the entire residence time of the experiment including post-shock equilibration and evacuation before sample collection. The influence of late-time gas dynamics on the effective residence time of morphological measurements and the associated uncertainty when comparing with modeled results is therefore of great interest \cite{ereminFormationCarbonNanoparticles2012}. To investigate this, a long-duration pressure trace ($\sim 35\,\mathrm{ms}$) was collected at a representative condition ($T_5$ 2238 K, $P_5$ = 4.5 atm) to measure expansion and compression waves after the nominal test time in the same facility. Particle formation in the shock tube was simulated using Omnisoot with a monodisperse particle dynamics configuration to reduce computational costs. The choice between a monodisperse and a sectional particle dynamics model was found to have little influence on predicted $\mathrm{CY}$ and $d_\mathrm{p}$.

Results indicate that $\mathrm{CY}$ (Fig.~\ref{fig:f_longt_ext}a) and $d_\mathrm{p}$ (Fig.~\ref{fig:f_longt_ext}b) reach a plateau within the first $5\,\mathrm{ms}$, while the decay and rise of $f_v$ follows the measured pressure trace as a direct result of changes in gas density, not mass growth. A compression wave returning after reflecting off the shock tube diaphragm section is seen in the pressure rise at $23\,\mathrm{ms}$ with no corresponding effect on $\mathrm{CY}$ or $d_\mathrm{p}$. The return of the reflected shock from the driver section endwall causes a second, stronger, compression at $28\,\mathrm{ms}$. The model predicts this pressure rise to influence CY on the order of $\Delta \mathrm{CY} = 0.007$ (4.34\%) and $d_\mathrm{p}$ on the order of $\Delta d_\mathrm{p} = 0.22\,\mathrm{nm}$ (1.42\%), relative to their $7.5\,\mathrm{ms}$ values. Model insights from Omnisoot indicate a large residual quantity of \ce{C2H2} is likely responsible for this additional growth enabled by a compression-driven temperature rise to $\sim$ 1400 K, briefly increasing HACA rates. Dilution, cooling, and a growing boundary layer act to continuously reduce the strength of subsequent compression wave and temperature rises, therefore having negligible later influence.

While these simulation results suggest that primary particle sizes and their internal nanostructure are not expected to evolve substantially between the end of the \textit{in situ} diagnostic window and sample recovery, they underscore secondary compression waves, facility dimensions, shock attenuation, and test-gas composition as important considerations in a complete picture of particle synthesis in shock tubes. Past work has rightly emphasized the need for consistent definitions of total observation time \cite{ereminExperimentalStudyMolecular2012, pitonExperimentalStudiesFormation2022}, but facility-specific long-time pressure histories and the associated uncertainty they bring to interpretations of microscopy results are not commonly reported. Such discussion is needed to help provide a practical basis for comparison of TEM-derived morphology to time resolved results and model predictions.

\subsection{Ex-situ carbon morphology and nanostructure}
\label{sec:exsitu_morphology}

\subsubsection{TEM images of synthesized particulate}
\label{sec:TEM_images}

\begin{figure*}[h]
    \centering
    \includegraphics[width=1.0\linewidth]{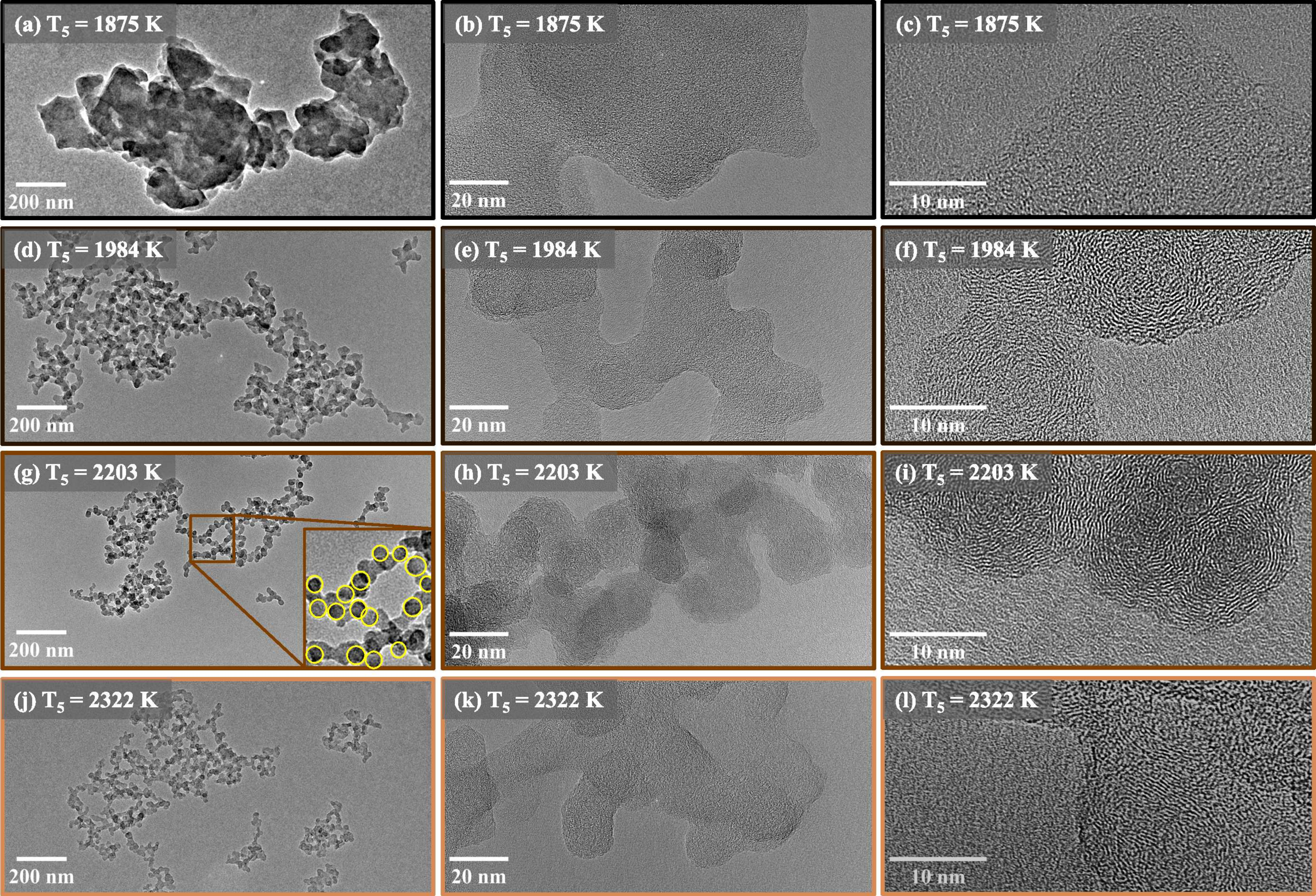}
    \caption{Overview of carbon particles collected at 1875 K (a–c), 1984 K (d–f), 2203 K (g–i), and 2322 K (j-l). For each temperature, panels show representative aggregates (left), primary-particle morphologies (middle), and high-resolution images highlighting internal structures (right). Panel (g) includes yellow circles demonstrating the manual measurements of primary particles done in ImageJ.}
    \label{fig:fig10_TEM_overview}
\end{figure*}

During CB and \ce{H2} production from \ce{CH4} pyrolysis, a better understanding of particle size and nanostructure supports the development of accurate model predictions beyond those of growth rates and yields. Figure~\ref{fig:fig10_TEM_overview} provides an overview of the synthesized carbon particles at representative low (a--c), intermediate (d--i), and high (j--l) post-reflected-shock temperatures. Below 1900~K (Figs.~\ref{fig:fig10_TEM_overview}a-c), non-spheroidal aggregates with indistinct primary particles formed. Above 1984 K, aggregate-scale morphology remained similar as temperature increased, possessing a fractal-like ramified structure composed of distinct primary particles. These aggregates are a product of the CB growth sequence in which a population of nascent particles forms first and subsequently evolves through coalescence, growth, carbonization and agglomeration. However, important differences in morphology emerge at the primary-particle and internal-structure scales as temperature changes.

At intermediate magnification (Figs.~\ref{fig:fig10_TEM_overview}b,e,h,k), the particles appear progressively smaller with increasing $T_5$; quantitative size statistics are reported in Section~\ref{sec:primary_particle_size_distributions} below. High-resolution panels further indicate that internal order evolves with temperature. Figure~\ref{fig:fig10_TEM_overview}i shows pronounced graphitic crystallites in the particles at 2203\,K, whereas the lower-temperature representative images (Figs.~\ref{fig:fig10_TEM_overview}c,f) show comparatively weaker fringe contrast, and the representative image at a higher temperature (2322~K) retains visible fringes that appears more heterogeneous but with less curvature (Fig.~\ref{fig:fig10_TEM_overview} l). These characteristics directly impact optical and chemical properties \cite{kelesidisSootLightAbsorption2019, vanderwalSootOxidation2003} and motivate quantitative statistical analysis of nanostructure statistics (Section~\ref{sec:carbon_particle_nanostructure_analysis}) to help connect time-resolved optical maturity (via $R(t)$) and the particle nanostructure recovered on TEM grids.

\subsubsection{Primary particle sizes and temperature trends}
\label{sec:primary_particle_size_distributions}

\begin{figure}[h]
    \centering
    \includegraphics[width=1.0\linewidth]{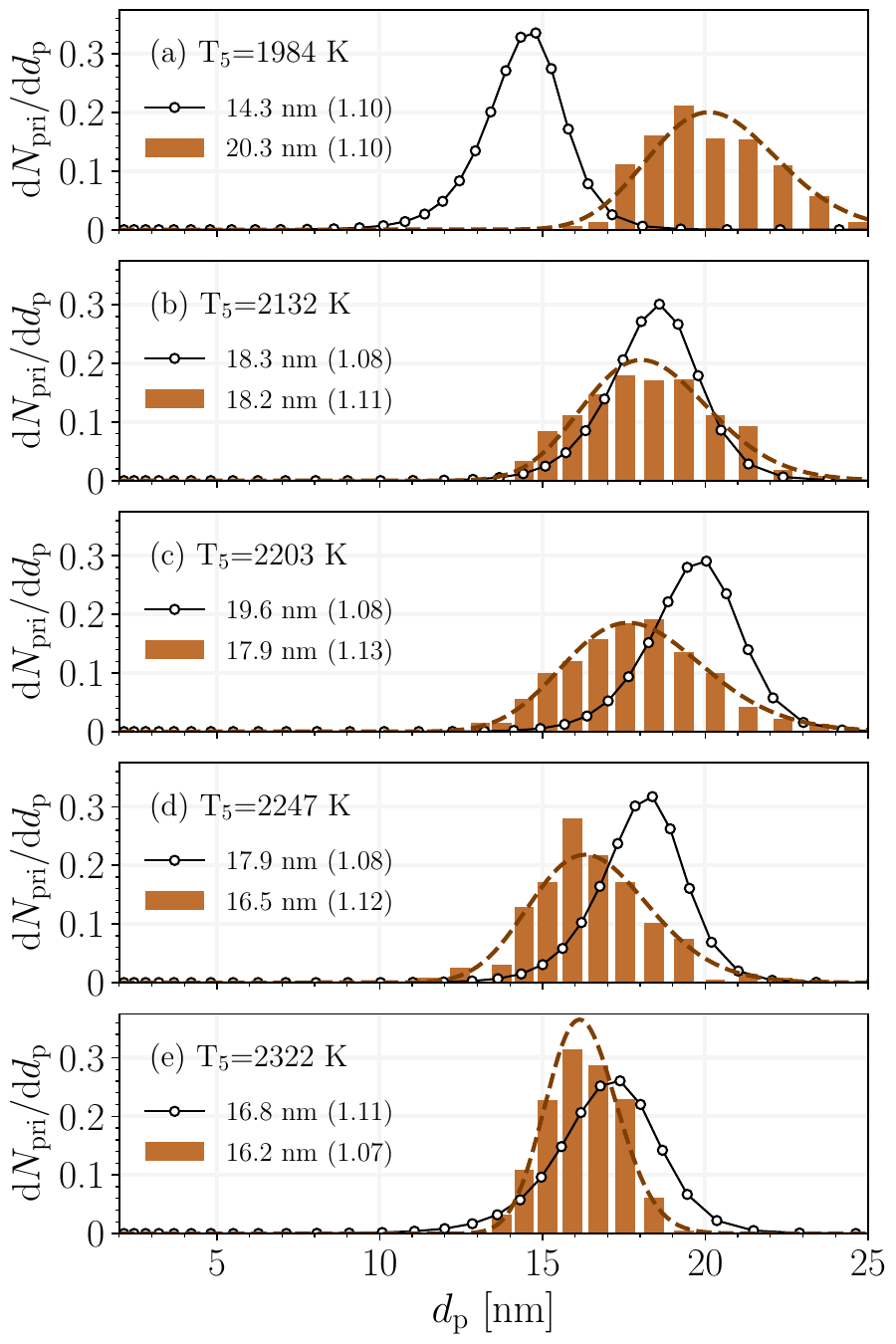}
    \caption{Primary particle size distributions measured from TEM analysis (brown histograms) at 1984, 2132, 2203, 2247, and 2322 K are well-fit by lognormal distributions with mean and standard deviation shown in the legend as $d_{p}$ $\mathrm{nm}$ ($\sigma_g$). Comparison with size distributions predicted by Omnisoot (white markers) at 7.5 ms indicate the model closely represents distribution widths ($\sigma_{g}$) but show a geometric mean not reflected in the measurements.}
    \label{fig:fig_PPSD}
\end{figure}

\begin{figure}[h]
    \centering
    \includegraphics[width=1.0\linewidth]{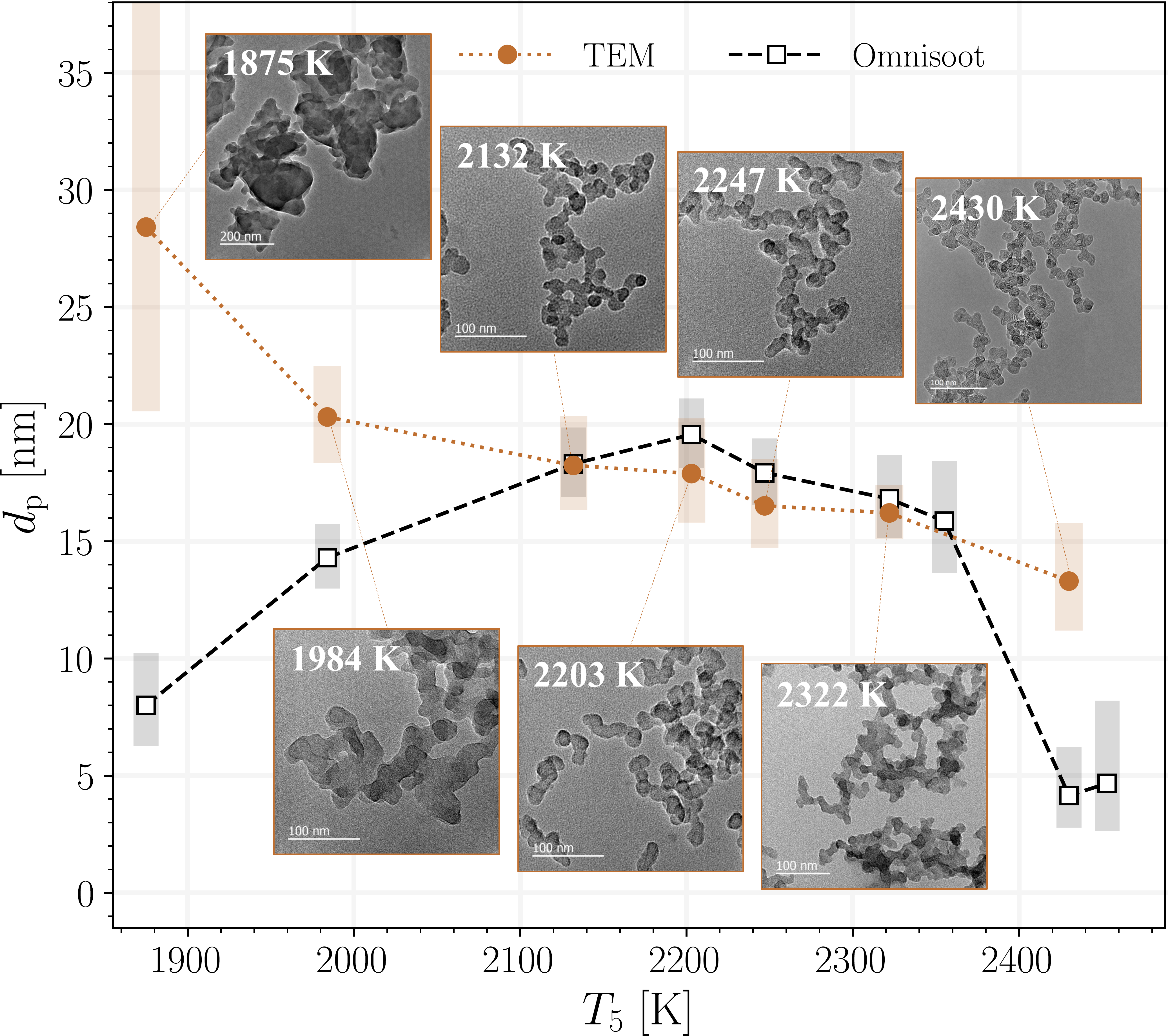}
    \caption{Measured and simulated geometric mean primary particle diameters, $d_{\mathrm{p}}$ reveal a higher $T_5$ decreases $d_\mathrm{p}$ towards 13.3 nm at 2430\,K while Omnisoot predictions agree well at mid temperatures, but show a bell-shaped trend, with under predictions at $2000\,\mathrm{K}\leq T_5 \leq 2350\,\mathrm{K}$. }
    \label{fig:fig_dp_vs_T5}
\end{figure}

Figure~\ref{fig:fig_PPSD} compares primary particle size distributions (PPSD) from \textasciitilde 200 manually measured primary particles per condition to predictions from Omnisoot. A PPSD was not calculated at 1875~K because of a lack of distinct primary particles to provide sufficient statistics. From 1950-2350\,K the measured distributions are well represented by lognormal fits with narrow widths ($ 1.07\leq\sigma_g \leq 1.13$) which suggests a confined time for inception followed by a homogeneous window of growth at each measurement condition. Predicted distribution widths from Omnisoot agree well with the measurements by describing inception with the E-bridge mechanism and growth via HACA and PAH adsorption, however above 2350\,K predicted size growth was insufficient to make a distribution comparison. Similarly narrow size distributions with means between 15 and 30~nm have also been reported using TEM from shock tube pyrolysis of other hydrocarbons including \ce{C2H2} \cite{knorreSootFormationPyrolysis1996}, \ce{C2H4}, and \ce{C6H6} \cite{bauerleSootFormationElevated1994}. 

Measurements of geometric mean primary particle diameter for \ce{CH4} pyrolysis clearly show a decreasing trend with $T_5$ as emphasized in Fig.~\ref{fig:fig_dp_vs_T5}. Measured $d_\mathrm{p}$ falls from 20.3~nm at 1984~K to 13.3~nm at 2430~K. Distinct primary particles were not formed at 1875~K, and thus the large geometric standard deviation error bars on $d_\mathrm{p}$ at 1875~K reflect the spread of approximate primary structures visible in TEM. Omnisoot does not reproduce the measured mean trend: it exhibits a bell-shaped mean that matches at intermediate temperatures, but underpredicts $d_\mathrm{p}$ on both ends. Omnisoot does capture a decrease in $d_\mathrm{p}$ at higher $T_5$ that results in good predictions between 2100 and 2350\,K by way of a carbonization mechanism that reduces particle surface reactivity in relation to modeled H/C ratio described in \cite{kholghyCoreShellInternal2016}. Detailed evaluation and optimization of Omnisoot parameters including carbonization for shock-driven CB synthesis is explored in a companion study currently in preparation by Adib et al. \cite{adibClark_ppr2}.

The large underprediction at 1984~K is an instructional result when taken together with the excellent agreement of time resolved $f_\mathrm{v}$ predictions from Fig.~\ref{fig:fig_fv}. It elucidates how a volume fraction history (or a late-time yield) can be well predicted, while distributing the condensed-phase carbon incorrectly between particle number and particle size. Morphological parameters like $d_\mathrm{p}$ thus provide a crucial constraint that can help reveal gaps in growth pathways for CB from \ce{CH4} pyrolysis over a given thermal history.

While previous shock tube pyrolysis studies have reported a similar decrease in $d_\mathrm{p}$ with $T_5$ for \ce{C2H4} \cite{deiuliisShockTubeBurnerThesis2009}, \ce{C6H6} \cite{pitonExperimentalStudiesFormation2022}, toluene (\ce{C7H8}), \cite{douceSootFormationHeavy2000,mathieuCharacterizationAdsorbedSpecies2007, pitonExperimentalStudiesFormation2022}, cyclopentene (\ce{C5H8}) \cite{pitonExperimentalStudiesFormation2022}, and heavier hydrocarbons \cite{douceSootFormationHeavy2000}, very few reports offer explanations for this relationship. The limited discussion that does exist suggest a depleted pool of PAH growth species \cite{mathieuCharacterizationAdsorbedSpecies2007}, or the increased production of heteroatoms and decreased interlayer spacing \cite{douceSootFormationHeavy2000} could be responsible for this behavior. However another explanation for smaller $d_\mathrm{p}$ in higher $T_5$ experiments can be considered in light of the multiwavelength extinction presented in this work. The spectral maturity ratio, $R(t)$, presented in Fig.~\ref{fig:fig_LEx}c shows that increasing $T_5$ has a drastic effect on its rate of decay, which is correlated with many components of maturity including a change in optical properties driven by carbonization, internal restructuring and an increasingly graphitic nanostructure (detailed in Section \ref{sec:carbon_particle_nanostructure_analysis}) \cite{gurentsovEffectSizeStructure2022}. This could explain the $d_\mathrm{p}$ decrease we observe with higher $T_5$ because the increased order driven by carbonization has been shown to lower surface reactivity by reducing the number of active sites \cite{Donnet1993_CBtextbook,vanderwalSootOxidation2003}, such that surface reactivity is driven down faster in high $T_5$ experiments. The data therefore suggest that the distribution of particles synthesized under high $T_5$ conditions mature and become less reactive sooner than at low $T_5$ and experience correspondingly less total surface growth as a result. Consequently, a larger proportion of carbon flux at higher $T_5$ is driven through inception pathways, resulting in smaller primary particles. Finally, the role of upstream gas chemistry in particle size trends should also be considered. LAS measurements of fuel conversion show at high temperatures, most fuel has been converted into \ce{C2H2} by 0.5 ms (Fig.~\ref{fig:fig_gas_yields}) and some consideration should be given to a rapid consumption of fuel via inception events at high $T_5$, leaving less total carbon for subsequent PAH formation as $T(t)$ drops, and therefore lower surface growth. 

Despite these self-consistencies, a number of shock synthesized carbon particle studies report a different influence of $T_5$ on $d_\mathrm{p}$. These reports include a bell curve (lower $d_\mathrm{p}$ at high and low $T_5$) \cite{kellererSootFormationShock1996a, starkeSootParticleSizing2001, ereminExperimentalStudyMolecular2012, ereminFormationCarbonNanoparticles2012}, or, from early investigations, no dependence \cite{bauerleSootFormationElevated1994, knorreSootFormationPyrolysis1996, deiuliisShockTubeBurnerThesis2009}. The latter trend tends to be discussed in conjunction with a high uncertainty band \cite{bauerleSootFormationElevated1994, knorreSootFormationPyrolysis1996}, or across a narrow $T_5$ range ($< 200$~K) \cite{deiuliisShockTubeBurnerThesis2009}. One possible reconciliation of the monotonically decreasing $d_\mathrm{p}$ trend discussed above with the $d_\mathrm{p}$ vs. $T_5$ bell curve reported elsewhere can be made by grouping the various studies based on measurement technique: studies that report a decreasing trend tend to have collected samples for TEM analysis, while studies that report bell curves extract size from time-resolved optical diganostics (LII or scattering). As is common for shock tube studies, the end of an optical experiment is typically defined by gas dynamics which disrupt the quiescent conditions required to make important assumptions of the gas medium. In the reported bell curves of $d_\mathrm{p}$ vs $T_5$, this chosen time in previous work is typically 1500~$\upmu$s. TEM measurements, meanwhile, include growth over the entire lifetime of the particle. At high $T_5$, particles may mature quickly due to rapid kinetics, thus the potential for continued growth after a short optical measurement time is low. As a result, the high-$T_5$ end of the ``bell curve'' maintains its decreasing trend during both \textit{in situ} optically- and \textit{ex situ} TEM-derived diameters. As discussed above and shown in Fig~\ref{fig:fig_LEx}c, however, the timescale for optical maturity is significantly lengthened at lower $T_5$, and the combination of slower kinetics and more active sites on lower $T_5$ particles in conjunction with slower fuel conversion could be one reason for particle growth beyond the reported diameter within a short optical diagnostic window. This would have the effect of \enquote{lifting up} the low-$T_5$ end of the \textit{in situ} bell curve. An example of such a comparison between LII at 1500~$\upmu$s and TEM results can be found in \cite{starkeSootParticleSizing2001}. 

\subsubsection{TEM image segmentation using a neural network}
\label{sec:tem_nn_segmentation}

\begin{figure}[h]
    \centering
    \includegraphics[width=1.0\linewidth]{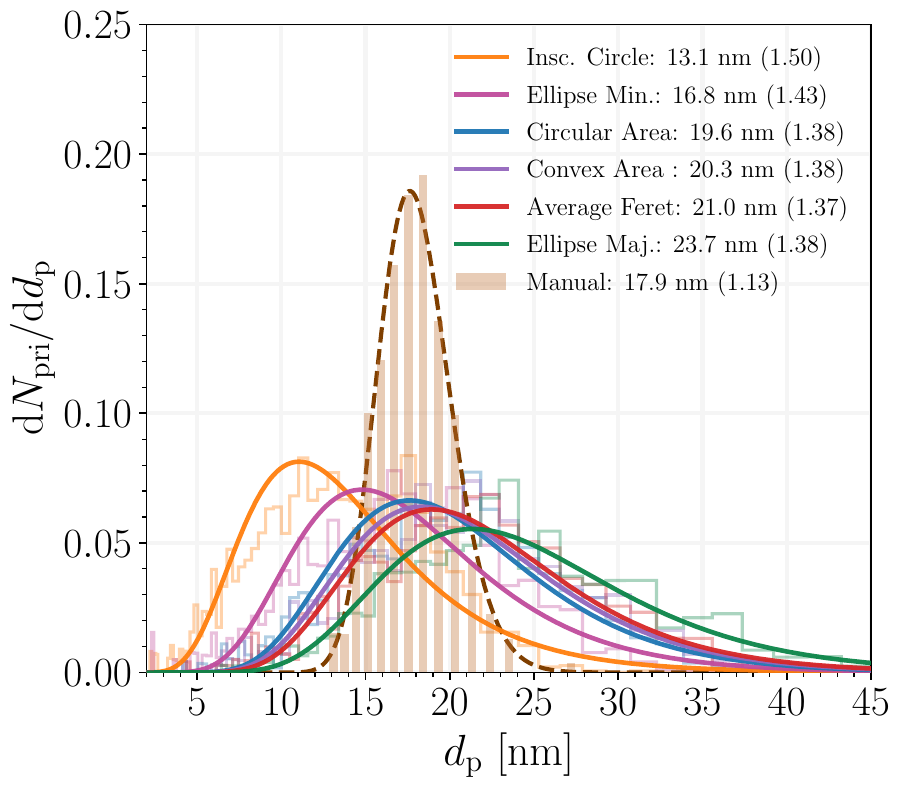}
    \caption{Raw primary particle size distributions and lognormal fits for 2203\,K returned from Cellpose-SAM segmentation in comparison with that produced by manual measurements in ImageJ with mean and standard deviation shown in the legend as $d_{p}$ $\mathrm{nm}$ ($\sigma_g$).}
    \label{fig:f_cellposeHist}
\end{figure}

Automated analysis of primary particle sizes addressed two limitations of manual ImageJ-based measurements: the labor required to generate statistically robust sample sizes and the difficulty of quantifying subjective decisions in particle selection and diameter measurement \cite{sipkensAtemsAnalysisTools2024}. In practice, collecting a 100-300 manual measurements per condition requires hours of researcher time across a handful of images. To scale primary particle size distributions (PPSDs) from hundreds of hand-labeled measurements to thousands of data points per condition, we applied a segmentation workflow using the Cellpose-SAM neural network (Section~\ref{methods_cellpose}). This automated form of $d_\mathrm{p}$ analysis resulted in primary-particle contours on the order of $10^3$ from the same image set in minutes. Figure~\ref{fig:f_cellposeMethods} illustrates a representative TEM input, the resulting contour overlays, and the motivation for the ensuing discussion: even for a single detected contour, multiple effective diameter definitions exist, each displaying a different sensitivity to variability in primary particles such as boundary irregularity, concavity, and elongation.

Figure~\ref{fig:f_cellposeHist} demonstrates that the choice of effective diameter metric strongly influences the returned PPSD even when segmentation contours are held fixed. For the same condition (2203~K), distributions computed from inscribed-circle, ellipse-axes, area-equivalent, convex-hull, and mean-Feret definitions exhibit offsets in both their modes and tails, spanning a broad range in mean diameter from 13.1 - 23.7 nm. This sensitivity is an important source of uncertainty in the automation of TEM-derived primary particle sizing. Using Cellpose-SAM it can be shown how definitions that respond to indentations (inscribed circle) differ from definitions that emphasize elongation or boundary extent such as the major ellipse axis or average Feret-based diameters. It is clear that the manual measurements form a noticeably tighter distribution, consistent with the human tendency to avoid ambiguity in densely overlapped regions and preferentially measure particles with clearly interpretable curvature. In contrast, the broader tails in the automated distributions reflect known failure modes in contour-based segmentation of carbon particle TEM images \cite{scottRapidAssessmentJet2024} including coalescence-driven contour merging and attempted segmentation of densely overlapped aggregate interiors (discussed below).

\begin{figure}[h]
    \centering
    \includegraphics[width=1.0\linewidth]{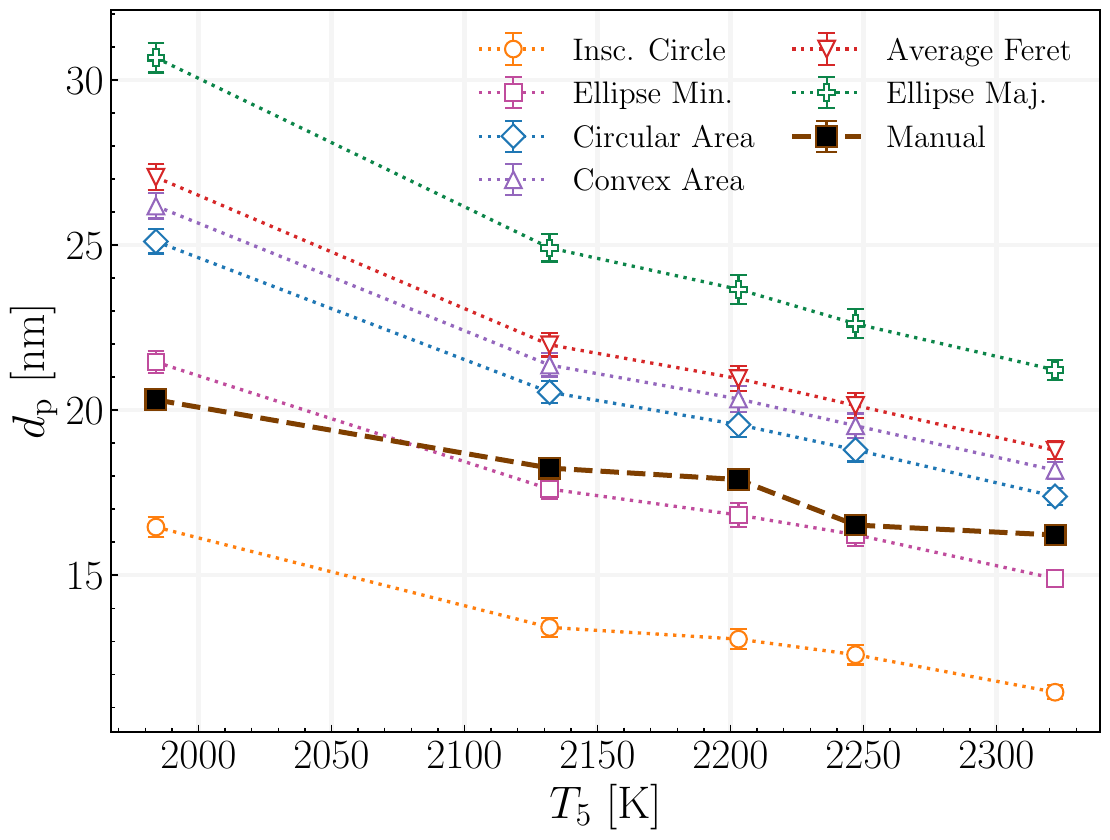}
    \caption{The inferred primary particle size from Cellpose-SAM segmentation of TEM images strongly depends on diameter metric chosen, but all metrics show the same trend with $T_5$. The minimum axis of the ellipse (Ellipse Min.) best matches the manual measurements from ImageJ in this work.}
    \label{fig:f_cellpose_dp_v_T5}
\end{figure}

Despite metric-dependent offsets, all automated diameter definitions recover the same temperature dependence for $d_\mathrm{p}$ as shown in Fig.~\ref{fig:f_cellpose_dp_v_T5}. The inferred geometric mean diameter decreases monotonically with increasing $T_5$ across all six metrics, indicating that this result is robust to reasonable choices of diameter definition. Comparison to manual measurements suggests how protocol-dependent bias can shape the ground truth $d_\mathrm{p}$. Although the manual workflow reports an average of the minimum and maximum Feret diameters for ellipses drawn to match perceived particle curvature, the automated mean-Feret metric is systematically larger than the manual dataset, whereas the ellipse-minor-axis metric aligns best with the manual trend seen in Fig.~\ref{fig:f_cellpose_dp_v_T5}. This agreement likely reflects a bias in aspects of the manual protocol where ellipse placement is guided by perceptible curvature in the presence of partially fused or ambiguous outlines. It should be noted that this does not suggest an ellipse-minor-axis definition is broadly best for automated segmentation.

There are many limitations of this automating primary particle sizing using pixel intensity \enquote{flows.} First, coalescence, necking, and the merging of outer graphitic layers can cause adjacent primary particles to appear as a single connected monomer, leading Cellpose-SAM to return a merged contour and biasing diameter estimates upward---particularly for boundary-extent metrics such as the mean Feret, which exceed our manual values. Second, the model may attempt to segment particles within the interior of densely overlapped aggregates where individual primaries are not visually separable and would be excluded by a human, which can also broaden PPSDs and distort distribution tails. The results presented here are therefore shown for raw segmentation outputs prior to post-filtering. Post-processing filters such as the exclusion of highly overlapped regions using ROI masks, circularity and aspect-ratio screening, and physically grounded size thresholds can be applied in future work to reduce the influence of these limitations and tighten PPSDs. Looking forward, the same segmentation outputs that enable rapid PPSD generation also provide a pathway to high-throughput aggregate descriptors (e.g., projected area, radius of gyration, and fractal dimension estimates), motivating the extension of this approach beyond primary particle sizing alone.

\subsubsection{Carbon particle nanostructure analysis}
\label{sec:carbon_particle_nanostructure_analysis}

\begin{figure}[!t]
    \centering
    \includegraphics[width=1.0\linewidth]{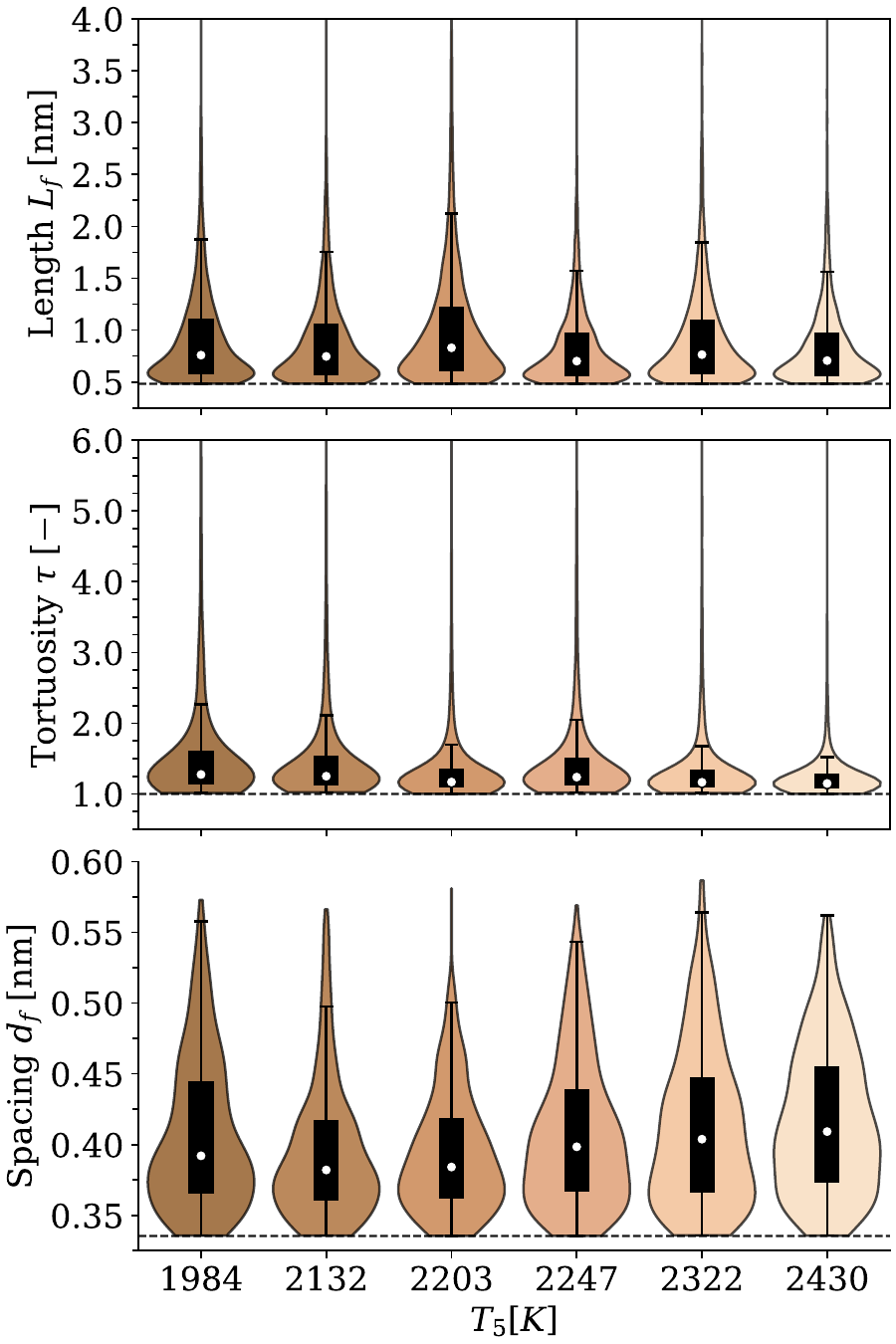}
    \caption{Distributions of HRTEM-derived nanostructure metrics versus $T_5$: (a) fringe length $L_\mathrm{a}$, (b) tortuosity $\tau$, and (c) interlayer spacing $d_\mathrm{f}$. For each $T_5$, violins show the kernel density of the measured population, the black box indicates the interquartile range (IQR) and white dots mark the median (which commmute with a log transform).}
    \label{fig:fig_fringe_T_trends}
\end{figure}

\begin{table*}[t]
\centering
\caption{Statistics of lattice-fringe metrics. At each $T_5$, $N=5000$ fringes are reported for $L_\mathrm{f}$ and $\tau$, and $N=500$ spacing samples for $d_\mathrm{f}$, of which the geometric mean (\(\mu_g\)), geometric standard deviation (\(\sigma_g\)), median, and the mode of the fitted lognormal distribution, are listed in the table.}
\label{tab:fringe_stats_compact_fullwidth}
\small
\setlength{\tabcolsep}{3.0pt}
\renewcommand{\arraystretch}{1.15}

\begin{tabular*}{\textwidth}{@{\extracolsep{\fill}}l rrrr rrrr rrrr}
\toprule
 & \multicolumn{4}{c}{$L_\mathrm{f}$ (nm)} & \multicolumn{4}{c}{$\tau$ (-)} & \multicolumn{4}{c}{$d_\mathrm{f}$ (nm)}\\
\cmidrule(lr){2-5}\cmidrule(lr){6-9}\cmidrule(lr){10-13}
$T_\mathrm{5}$ (K) & $\mu_\mathrm{g}$ & $\sigma_\mathrm{g}$ & Median & Mode & $\mu_\mathrm{g}$ & $\sigma_\mathrm{g}$ & Median & Mode & $\mu_\mathrm{g}$ & $\sigma_\mathrm{g}$ & Median & Mode\\
\midrule
1984 & 0.832 & 1.523 & 0.759 & 0.537 & 1.47 & 1.48 & 1.27 & 1.11 & 0.405 & 1.135 & 0.392 & 0.369\\
2132 & 0.812 & 1.506 & 0.746 & 0.533 & 1.42 & 1.42 & 1.25 & 1.10 & 0.392 & 1.119 & 0.382 & 0.361\\
2203 & 0.903 & 1.595 & 0.830 & 0.557 & 1.30 & 1.35 & 1.17 & 1.08 & 0.392 & 1.108 & 0.384 & 0.367\\
2247 & 0.767 & 1.451 & 0.701 & 0.529 & 1.40 & 1.42 & 1.24 & 1.10 & 0.405 & 1.129 & 0.398 & 0.373\\
2322 & 0.830 & 1.524 & 0.764 & 0.539 & 1.30 & 1.35 & 1.16 & 1.08 & 0.409 & 1.139 & 0.404 & 0.375\\
2430 & 0.774 & 1.455 & 0.706 & 0.535 & 1.23 & 1.25 & 1.15 & 1.09 & 0.414 & 1.130 & 0.409 & 0.391\\
\bottomrule
\end{tabular*}
\end{table*}

HRTEM indicates that internal organization changes markedly across $T_5$. The lowest-$T_5$ sample (1875~K) shows weak or absent lattice-fringe contrast (Fig.~\ref{fig:fig10_TEM_overview}c). At and above 1984~K, concentric ``onion-like'' fringe patterns become apparent, consistent with canonical descriptions of carbonization, in which disordered structures can progressively organize into continous layers \cite{kholghyCoreShellInternal2016, boteroInternalStructureSoot2019}. At 1984~K (Fig.\ref{fig:fig10_TEM_overview}f), fringes are present but diffuse, with partially developed shells and amorphous contrast that persists both within the particle interior and between stacks. At 2203~K (Fig.~\ref{fig:fig10_TEM_overview}i), fringe contrast is strongest and most spatially continuous, stacks are readily resolved and frequently trace smooth shell-like curvature around the primary particle periphery. At 2322~K (Fig.~\ref{fig:fig10_TEM_overview}l), fringe patterns remain apparent, yet appear to curve less around primary-particle boundaries and show less uniformity.

Quantitative and reproducible analysis of nanostructure from HRTEM across 1984--2430~K used the fixed-operator workflow detailed in Section~\ref{methods_nanostructure}. Lognormal distributions of $L_\mathrm{f}$, $\tau$, and $d_\mathrm{f}$ were calculated (accounting for the hard lower bounds imposed by the analysis definitions and acceptance criteria) and summarized in Table~\ref{tab:fringe_stats_compact_fullwidth}. The spread of distribution densities is reported in Fig.~\ref{fig:fig_fringe_T_trends}. Raw distributions at each $T_5$ can be found in Section~\ref{SM:nanoProcessing} of the S.M. The extracted metrics show weak temperature dependence relative to the qualitative contrast in Fig.~\ref{fig:fig10_TEM_overview}: median $L_\mathrm{f}$ is highest at 2203~K (0.830~nm), $\tau$ exhibits only a weak minimum at higher $T_5$, and $d_\mathrm{f}$ varies over a narrow range (geometric mean 0.392-0.414~nm). Notably, an increase in order that peaks at intermediate temperatures, before becoming more fragmented as $T_5$ increases further is strikingly similar to the behavior reported in toluene pyrolysis \cite{mathieuCharacterizationAdsorbedSpecies2007}.

The limitations of extracting such quantitative trends merit some discussion. First, the shorter $L_\mathrm{f}$ at 2247~K does not align with the qualitative progression from visual inspections, (where it does not appear significantly less ordered, see Section~\ref{SM:nanoProcessing} in S.M.), and the distributions of $d_\mathrm{f}$ are compressed by admissible-pair criteria (i.e., spacing is only computed for near-parallel, physically plausible separations) which can reduce sensitivity to modest structural shifts and make differences difficult to distinguish. Prior work has shown that different nanostructure metrics have varying discriminatory power, and that median-based summaries are often preferred for skewed fringe statistics  \cite{yehliuDevelopmentHRTEMImage2011,boteroInternalStructureSoot2019}. More fundamentally, this automated extraction relies on visible lattice contrast; if the contrast or the visibility of fringes is spatially heterogeneous, extracted fringes will preferentially sample certain subregions more accurately rather than reliably representing the full particle cross-section. This disconnect between visual impressions and quantified distributions should therefore be treated as a methodological uncertainty rather than evidence that ordering is invariant with temperature.

Nanostructure also lends an additional axis on which to interpret multiwavelength extinction and primary-particle size trends (Sections~\ref{sec:multiwavelength_LEx} and \ref{sec:primary_particle_size_distributions}). The regime with the clearest stacked fringes by visual inspection (near and above 2200~K) coincides with rapid decay of the optical maturity ratio $R(t)$ (Fig.\ref{fig:fig_LEx}c), consistent with literature interpretations linking increased sp$^2$ hybridization to a reduced optical band gap and an absorption response that is more independent of wavelength \cite{russoOpticalBandGap2020, hagenCarbonNanostructureReactivity2021}. Conversely, disordered fringe patterns at 1875~K align with measurements of limited optical maturity and prior observations that lower-temperature pyrolysis can retain more organic material \cite{mathieuCharacterizationAdsorbedSpecies2007}. Furthermore, increased graphitic-layer continuity is commonly associated with fewer reactive edge sites and reduced surface reactivity, supporting a structural basis for the observed decrease in primary-particle size with increasing $T_5$. At the same time, the coexistence of ordered and disordered regions within a primary particle supports a heterogeneous maturity framework in which surface and bulk ordering can evolve differently \cite{johanssonEvolutionMaturityLevels2017}. Future work should consider region-resolved HRTEM analysis (shell vs. core, or high- vs. low-contrast regions) to examine how heterogeneity influences automated fringe statistics, and to tighten the linkage between nanostructure, optical maturity, and temperature-dependent growth behavior in pyrolytic synthesis processes.

\section{Conclusions}
\label{conclusion}

This work presents a combined experimental and modeling study of the gas-phase chemistry, time-resolved particle formation, and ex-situ morphology relevant to CB formation from the thermal pyrolysis of \ce{CH4}. Initial carbon flux through small hydrocarbons provided by LAS measurements of CH$_4$, C$_2$H$_4$ and C$_2$H$_2$ in a shock tube reveals strong agreement between FFCM-2, CRECK, and CALTECH kinetic models across the temperature range, while wide variations in predicted PAH concentrations persist, motivating future work to probe PAH kinetics in shock tube experiments. Gas speciation also informed the choice of the CRECK kinetic model to describe gas chemistry for particle inception and growth simulations in Omnisoot. Multiwavelength light extinction measured particle $\tau_\mathrm{ind}$, $f_v$, CY, and millisecond-scale evolution of optical maturity. The more rapid decay of the spectral maturity ratio $R(t)$ at higher $T_5$ is correlated with less mass growth over the residence time as found in $d_\mathrm{p}$ measurements from TEM images. Stringent model evaluation is provided by these concurrent measurements: Omnisoot reproduces the magnitude and pressure-driven decay of $f_\mathrm{v}$, but shows systematic errors at both temperature extremes of the particle formation \enquote{bell curve.} Predicted inception behavior seen during $\tau_\mathrm{ind}$ measurements aligns closest with the delayed extinction measured at 1064 $\mathrm{nm}$, but becomes increasingly delayed and non-Arrhenius above $\sim2200\ \mathrm{K}$. The reported model comparisons point to deficiencies in the temperature scaling of PAH precursor chemistry or inception pathways in CH$_4$ pyrolysis, with current measurements providing opportunities to improve model agreement. Finally, simulations of long-time shock tube gas dynamics indicate that subsequent compressions can increase $\mathrm{CY}$ on the order of $5$\% and $d_\mathrm{p}$ by $\leq 2\%$ after the \textit{in situ} diagnostic window at the presented conditions, improving the understanding of limitations and uncertainties when making comparison of TEM-derived morphology to time-resolved measurements.

TEM measurements revealed narrow lognormal PPSDs and a monotonic decrease in $d_\mathrm{p}$ with increasing $T_5$ from 20.3 $\mathrm{nm}$ at $1984\ \mathrm{K}$ to $ 13.3\ \mathrm{nm}$ at $2430\ \mathrm{K}$, indicating increased inception flux relative to growth pathways, which is correlated with rapidly maturing particles at higher $T_5$. Omnisoot predicted particle size and $T_5$ trends well between 2100\,K and 2350\,K, but comparison of combined $f_\mathrm{v}$ and $d_\mathrm{p}$ measurements with Omnisoot also highlight that agreement in only one metric can mask incorrect partitioning of condensed carbon between particle mass and morphology. The automation of $d_\mathrm{p}$ analysis using Cellpose-SAM enabled more extensive, efficient, and repeatable particle characterization, while highlighting the bias introduced when selecting a diameter metric and showing potential for advanced CB morphological quantification in the future. Finally, HRTEM indicated the most distinct, contiguous nanostructure near $2200\ \mathrm{K}$ (coinciding with peak CY), while disordered below $1900\ \mathrm{K}$. More work is needed to correlate quantitative fringe metrics to $d_\mathrm{p}$ and high-temperature $R(t)$ .

Collectively, the present study combines time resolved measurements of gas chemistry and particle formation with \textit{ex situ} size and nanostructure to deepen understandings and improve the fidelity of modeling tools that describe the phenomena driving CB and H$_2$ production from CH$_4$ pyrolysis. The results identify PAH precursor concentrations, low $T_5$ surface growth, inception chemistry at high $T_5$, and the coupling of maturity-dependent optical properties and nanostructure with mass growth rates as key targets for improving predictive models of CB formation from \ce{CH4} pyrolysis.

\section{Acknowledgments}
This research was supported by Monolith Inc. with contract number AW945796 managed by Dr. Enoch Dames and Dr. Shruthi Dasappa. Part of this work was supported by the National Science Foundation under award ECCS-2026822. Work at Carleton University was supported by the Canada Research Chairs Program (Grant CRC-2019-232527), the Natural Sciences and Engineering Research Council of Canada (NSERC) through a Discovery Grant (RGPIN-
2019-06330), and MITACS (IT42881). Part of the work was also performed at nano@stanford RRID:SCR\_026695 with the support of Andrew Barnum, Paul Hart, and Pinaki Mukherjee.

The authors would also like to thank Dr. Hai Wang and Dr. Adam Boies for their feedback, Dr. Luke Zaczek for essential input to the experimental planning and execution, Dr. Lingzhi (Jackie) Zheng for image segmentation discussions, as well as Dr. Chuyu Wei, Colin Madaus, and Naman Mishra for their support during data collection.

\bibliographystyle{elsarticle-num} 
\bibliography{references}

\baselineskip 9pt
\clearpage

\newcommand{\beginsupplement}{%
	\setcounter{figure}{0}%
	\renewcommand{\thefigure}{S\arabic{figure}}%
	\setcounter{table}{0}%
	\renewcommand{\thetable}{S\arabic{table}}%
	\setcounter{equation}{0}%
	\renewcommand{\theequation}{S\arabic{equation}}%
	\setcounter{section}{0}%
	\renewcommand{\thesection}{S\arabic{section}}%
	\setcounter{page}{1}%
	\renewcommand{\theHfigure}{S\arabic{figure}}%
	\renewcommand{\theHtable}{S\arabic{table}}%
	\renewcommand{\theHequation}{S\arabic{equation}}%
	\renewcommand{\theHsection}{S\arabic{section}}%
}
\clearpage
\onecolumn
\setcounter{page}{1}
\beginsupplement
\normalsize
\baselineskip 12pt
\setlength{\parindent}{0pt}
\setlength{\parskip}{0.6em}

\begin{center}
{\Large \textbf{Supplementary Material}}\\[0.4cm]
{\large Carbon black and hydrogen production from methane pyrolysis: measured and modeled insights from integrated gas and particle diagnostics in shock tubes}\\[0.6cm]

Gibson Clark$^{a,*}$, Mo Adib$^{b}$, Chengze Li$^{a}$, Taylor M. Rault$^{a}$,\\Jesse W. Streicher$^{a}$, Enoch Dames$^{c}$, Reza Kholghy$^{b}$, Ronald K. Hanson$^{a}$\\[0.4cm]

$^{a}$Department of Mechanical Engineering, Stanford University, Stanford, CA 94305, USA\\
$^{b}$Department of Mechanical Engineering, Carleton University, Ottawa, ON K1S 5B6, Canada\\
$^{c}$Monolith Materials, San Carlos, CA 94070, USA

$^{*}$Corresponding author: gibsonc@stanford.edu

\end{center}

\section{New measurements of absorption coefficients in \ce{CH4} pyrolysis}
\label{SM:absorption_coefficients}


New absorption coefficient measurements at 2998~nm and 3366~nm were made to enable laser absorption spectroscopy results with reduced spectral interference between major pyrolysis products during high fuel-loading experiments ($\chi_{\mathrm{CH}_4} \geq 5\%$). All measurements were made in a bath gas of Argon

\begin{figure}[htb]
    \centering
    \includegraphics[width=1.0\linewidth]{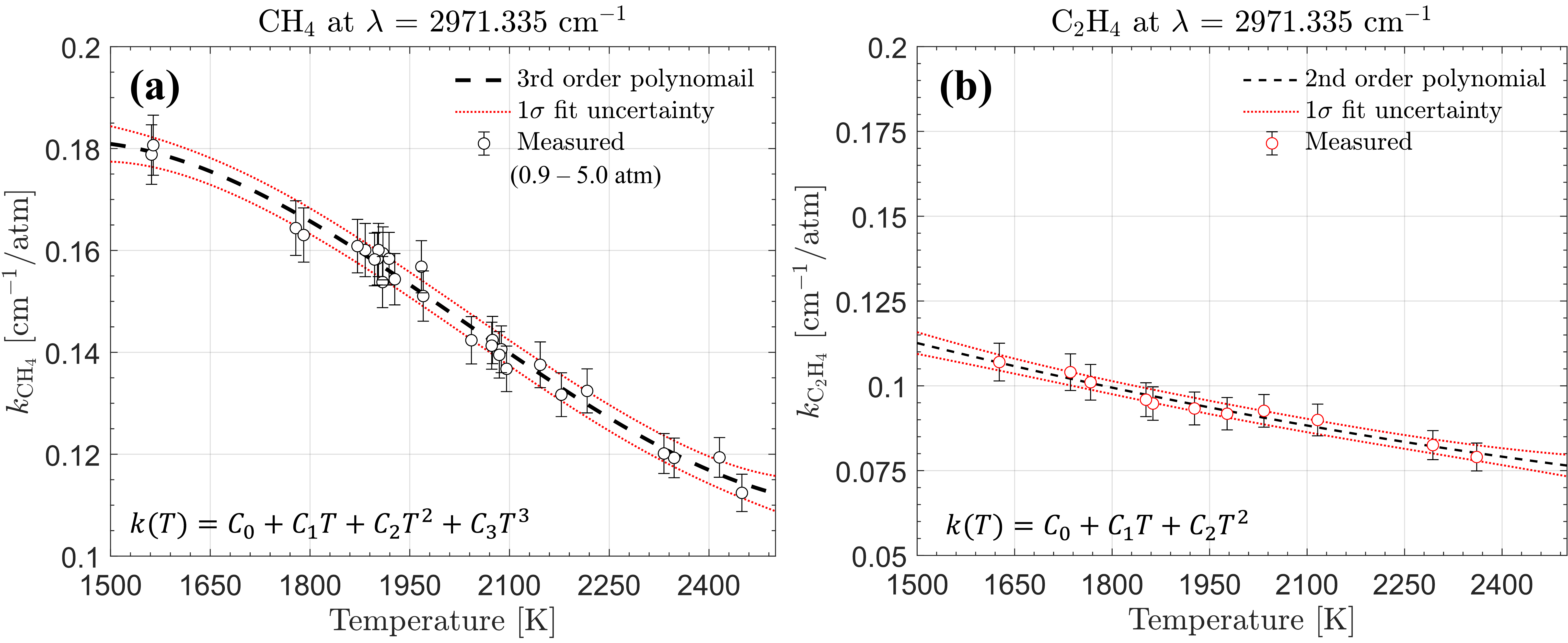}
    \caption{\footnotesize Temperature-dependent absorption coefficient measurements at 3366~nm. (a) Polynomial correlation for \ce{CH4} used in this work, based on measurements spanning 0.9--5~atm. (b) Measured \ce{C2H4} absorption coefficients at the same wavelength to quantify spectral interference from \ce{C2H4}. Measured error bars represent the addition in quadrature of uncertainties in $T_5$, $P_5$, and time-zero absorbance, while the plotted fit uncertainty shows a 1-standard deviation confidence interval of the fit.}
    \label{fig:f_supp_k3365nm}
\end{figure}

Figure~\ref{fig:f_supp_k3365nm} summarizes the resulting temperature-dependent correlations at 3366~nm. The measurements of \ce{CH4} absorption coefficients in Fig.~\ref{fig:f_supp_k3365nm}a are the result of a cluster of high energy C-H stretch vibrations in the P-branch which give rise to broad, high-temperature spectral features. The nature of many blended lineshapes thus render the effects of pressure dependent broadening between 0.9 and 5.0 atm negligible. The absorption coefficient is expressed as $k(T)$ in units of $\mathrm{atm}^{-1}\,\mathrm{cm}^{-1}$ and $T$ is in K:

\begin{equation}
k_{\ce{CH4},3366\mathrm{nm}}(T)
= 8.467\times10^{-11}T^{3}
      -5.169\times10^{-7}T^{2}
 +9.618\times10^{-4}T
      -3.845\times10^{-1}
\label{eq:k_ch4_work}
\end{equation}

Figure~\ref{fig:f_supp_k3365nm}b shows corresponding measurements of \ce{C2H4} absorption coefficients at the same wavelength, obtained to quantify spectral interference from \ce{C2H4} where Eq.~(\ref{eq:k_c2h4}) is used to estimate the absorbance contribution of \ce{C2H4} at 3366~nm using a known concentration through the Beer-Lambert law. Because \ce{C2H4} has a smaller absorption coefficient than \ce{CH4}, and because it is present at much lower mole fractions than \ce{CH4}, the inferred \ce{C2H4} interference typically remained below 2\% of the measured signal and peaked on the order of 5\% when \ce{C2H4} formation coincided with lower \ce{CH4} concentrations. Uncertainties in these correlations of of $\approx$ 7\% were calculated via addition in quadrature of 2\% uncertainty in $T_5$ and $P_5$, in addition to experimental uncertainty in the time-zero absorbance measured to calculate $k(T)$.

\begin{equation}
k_{\ce{C2H4},3366\mathrm{nm}}(T)
= 1.101\times10^{-8}T^{2}
-8.017\times10^{-5}T+2.081\times10^{-1}
\label{eq:k_c2h4}
\end{equation}

\begin{figure}[h]
    \centering
    \includegraphics[width=0.5\linewidth]{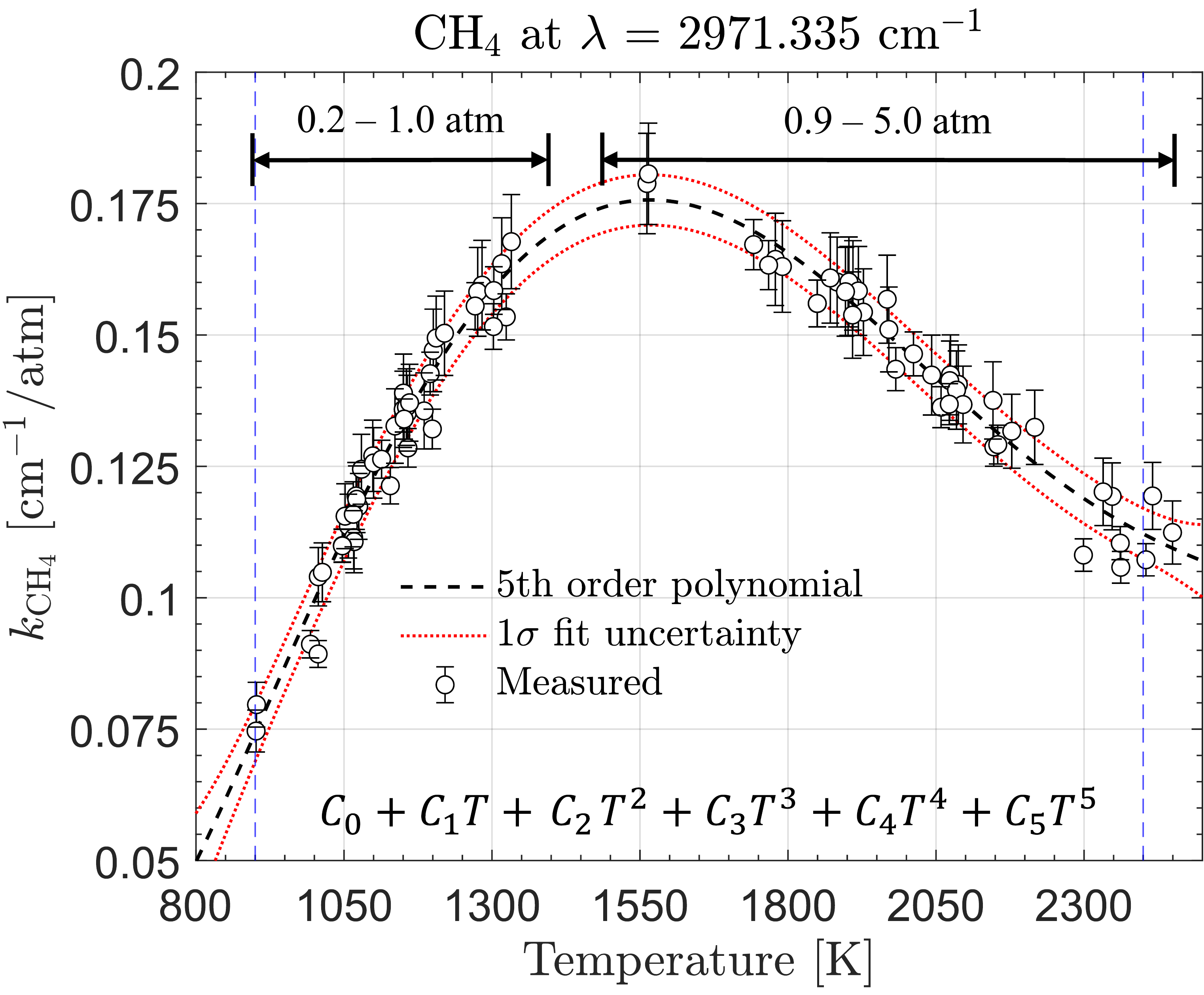}
    \caption{\footnotesize Temperature-dependent absorption coefficient measurements of \ce{CH4} at 3366~nm across an extended temperature range for applications in low-temperature pyrolysis studies; this correlation is valid between 900 and 2400 K. Measured error bars represent uncertainties in $T_5$, $P_5$, and time-zero absorbance and a 1-standard deviation confidence interval of the fit is shown in dotted red.}
    \label{fig:f_supp_k3365nm_allT}
\end{figure}

Figure \ref{fig:f_supp_k3365nm_allT} extends the \ce{CH4} temperature range for future use of this diagnostic in lower-temperature pyrolysis studies; these  measurements below 1400~K were obtained between 0.2--1.0~atm. The broad temperature range fit is given as:

\begin{equation}
k_{\ce{CH4},3366\mathrm{nm\,ext}}(T)
=
-7.19\times10^{-17}T^{5}
+6.64\times10^{-13}T^{4} 
-2.28\times10^{-9}T^{3} 
+3.49\times10^{-6}T^{2}
-2.19\times10^{-3}T
+4.85\times10^{-1}
\label{eq:k_ch4_extended}
\end{equation}
\vspace{1pt}

This extended correlation should be applied with caution below 900~K or above 2400~K, limits indicated by the vertical dashed lines. At elevated temperature, rapid \ce{CH4} decomposition increases the uncertainty in the inferred time-zero absorbance and therefore in the derived absorption coefficient.

\par
\begin{figure}[h]
    \centering
    \includegraphics[width=0.5\linewidth]{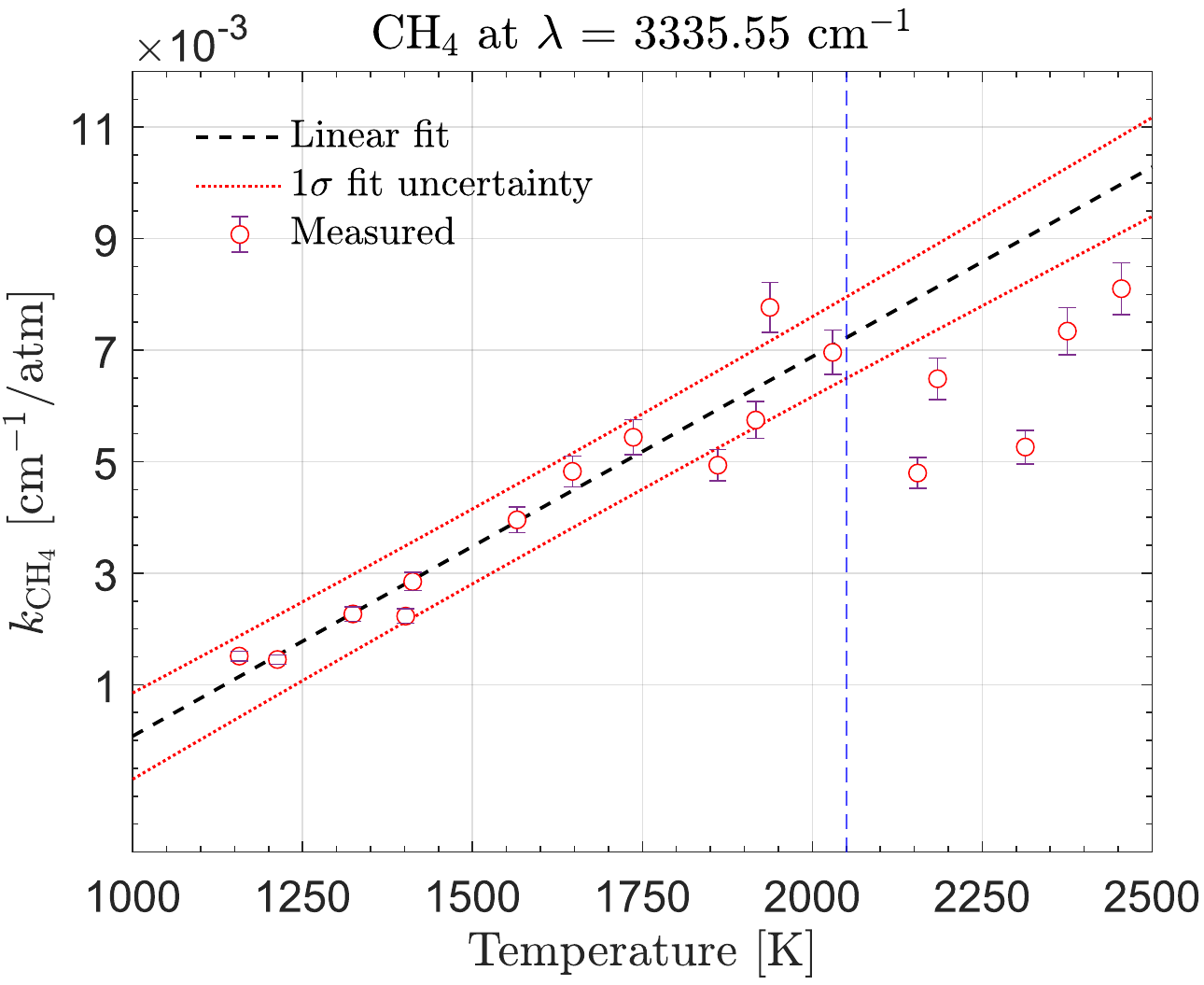}
    \caption{\footnotesize Temperature-dependent \ce{CH4} absorption coefficient measurements at 2998~nm used to quantify spectral interference during \ce{C2H2} measurements. Because \ce{CH4} absorbs at 2998~nm, its contribution to the measured absorbance was estimated and removed when inferring \ce{C2H2} mole fraction time histories. This correction was most important at early times and under conditions where \ce{CH4} remained large while \ce{C2H2} was still small.}
    \label{fig:f_supp_k2998nm_ch4}
\end{figure}

 Additional measurements were also performed at 2998~nm (3335.555 $\mathrm{cm}^{-1}$) to quantify spectral interference from \ce{CH4} on the \ce{C2H2} diagnostic wavelength \cite{stranicLaserAbsorptionDiagnostic2014}. Interference from \ce{CH4} is exacerbated at high fuel loadings and can contribute to the measured absorbance at 2998 nm, particularly at early times when \ce{CH4} concentrations remain high but \ce{C2H2} concentrations are low. As a result, neglecting \ce{CH4} interference will bias inferred \ce{C2H2} mole fractions upward under some conditions. Figure~\ref{fig:f_supp_k2998nm_ch4} provides temperature-dependent \ce{CH4} absorption coefficient measurements at 2998~nm to estimate this interference. These measurements were used together with the measured \ce{CH4} mole fraction time history to calculate the \ce{CH4} contribution to the total absorbance at 2998~nm and isolate the \ce{C2H2} signal. The importance of this correction was greatest at early times and under lower-temperature conditions, where \ce{CH4} conversion was incomplete and \ce{C2H2} mole fractions were comparatively small. The temperature-dependent \ce{CH4} absorption coefficient correlation at 2998~nm is given by

\begin{equation}
k_{\ce{CH4},\,2998}(T)
=
6.805\times10^{-6}T
-6.727\times10^{-3}
\label{eq:app_k_ch4_2998}
\end{equation}
\vspace{1pt}

where again, $k(T)$ is in units of $\mathrm{atm}^{-1}\,\mathrm{cm}^{-1}$ and $T$ is in K. The correlation in Eq.~(\ref{eq:app_k_ch4_2998}) and plotted in Fig.~\ref{fig:f_supp_k2998nm_ch4} is fit only to temperatures below 2050~K, as indicated by the dashed vertical line, due to very low SNR and thus high uncertainty in time-zero absorbance at higher $T_5$ at this wavelength. The \ce{CH4} interference in \ce{C2H2} measurements was therefore measured to account for 0.001 to 0.004 in decimal mole fraction space, ranging between 2-10\% of the true \ce{C2H2} concentration formed during the test-time, depending on $T_5$.

\FloatBarrier 

\section{Preprocessing for image segmentation}
\label{SM:cellPose_workflow}

TEM images were preprocessed before Cellpose-SAM segmentation to standardize the analysis region and improve contrast for contour detection. Images with multiple channels were first converted to grayscale. A fixed square crop was applied to the image to exclude the scale bar and edge artifacts. Within this cropped image, three-class multi-Otsu thresholding was used to identify dark particle regions, intermediate intensities, and bright background pixels. Pixels above the upper Otsu threshold were reassigned to the mean intensity of the bright class to flatten background variation, and pixels below two-thirds of the lower Otsu threshold were clamped to that value to reduce spurious segmentation in dark, highly overlapped regions. Finally, intensities were clipped to the 1st and 99.5th percentiles and rescaled to an 8-bit range prior to Cellpose-SAM inference.

\begin{figure}[h]
    \centering
    \includegraphics[width=0.77\linewidth]{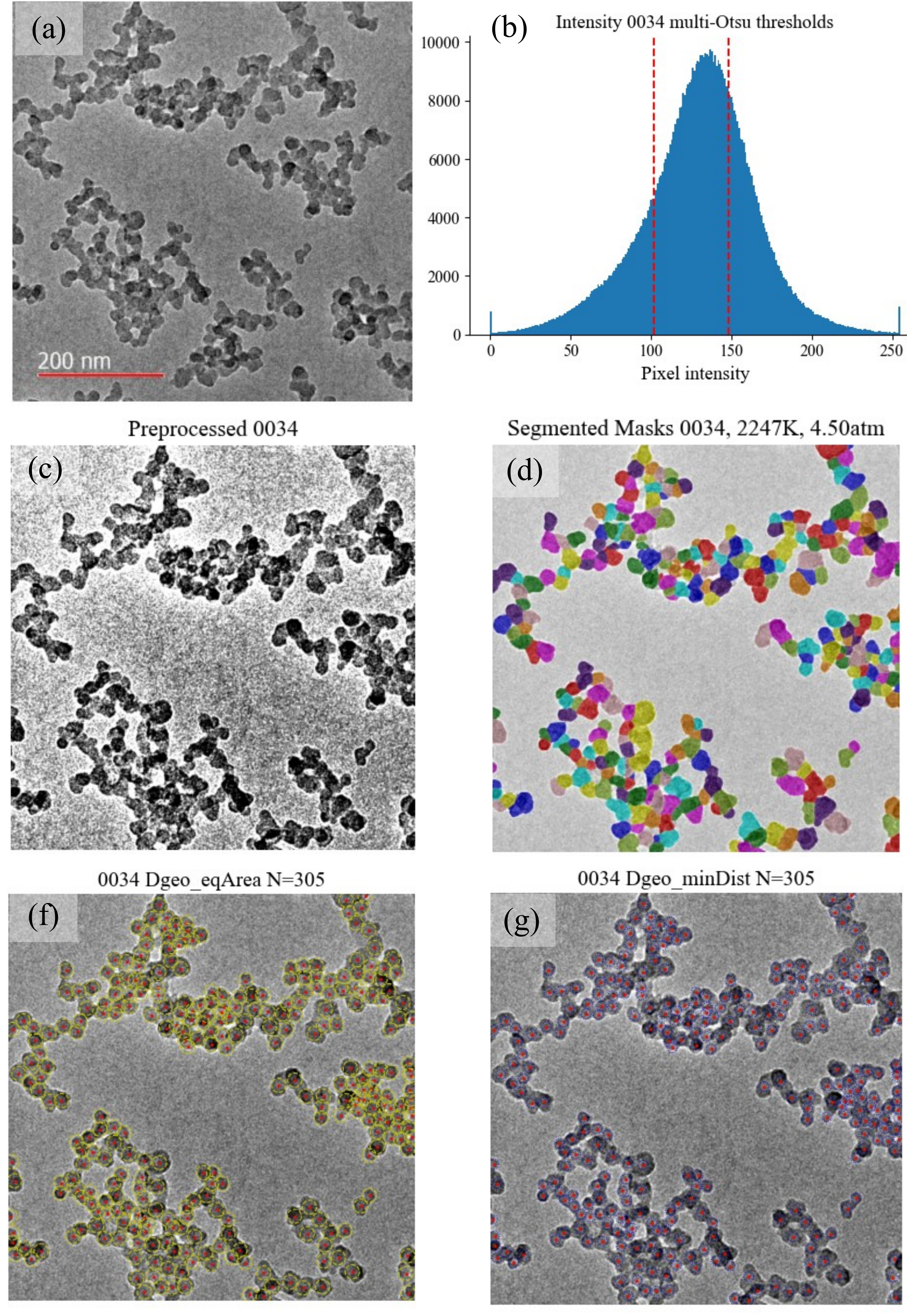}
    \caption{\footnotesize Preprocessing and segmentation workflow for a TEM image at $2247~\mathrm{K}$ and $4.5~\mathrm{atm}$. (a) Raw TEM image with scale bar detection. (b) Histogram showing multi-Otsu thresholds (red dashed lines). (c) Preprocessed image after background flattening, intensity clamping, and normalization. (d) Segmentation map showing individual contour masks. (f,g) Overlay of geometric mean equivalent-area-circle and minimum-distance diameters from the same contour set.}
    \label{fig:supp_cellPoseSAM}
\end{figure}

An example of this preprocessing workflow is shown in Fig.~\ref{fig:supp_cellPoseSAM}. The raw TEM image and scale bar detection are shown in Fig.~\ref{fig:supp_cellPoseSAM}a, while the multi-Otsu intensity histogram and threshold values are shown in Fig.~\ref{fig:supp_cellPoseSAM}b. The resulting preprocessed image is shown in Fig.~\ref{fig:supp_cellPoseSAM}c, and the corresponding Cellpose-SAM segmentation map is shown in Fig.~\ref{fig:supp_cellPoseSAM}d. After segmentation, extracted particle contours were used to compute geometric statistics for multiple diameter definitions, including equivalent-area diameter and minimum-distance-based diameter, as illustrated in Fig.~\ref{fig:supp_cellPoseSAM}f--g. In addition to diameter-based measurements, the same contour set can be filtered or analyzed using other descriptors such as particle area, circularity, aspect ratio, boundary curvature, and other shape-quality criteria.

Although the preprocessed image appears visually less natural than the raw TEM image, the grayscale shading, smooth background texture, and local contrast cues that help intuit where one particle ends and another begins required the above transformations to regularize contrast for reliable Cellpose-SAM inference.

\FloatBarrier 

\section{Automated nanostructure parameter selection}
\label{SM:nanoProcessing}

Fringe-extraction parameters (e.g., smoothing strength and kernel size, structuring-element size, and binarization thresholding factor) were tuned using representative images from each condition. For each trial parameter set, the resulting binary image and skeletonized image were first visually examined, and eventually extracted fringes were overlaid on the original micrographs for qualitative validation (Fig.~\ref{fig:fig_fringe_method}c), and parameters were iteratively adjusted to suppress common failure modes (spurious noise fringes, over-fragmentation, and excessive branching) while preserving visually apparent lattice features. The final operator set was then held fixed for all images to ensure consistent comparisons across conditions.

Representative HRTEM images at each $T_5$ are shown below in Figs.~\ref{fig:fig_HRTEM_supp} a-f, and histograms of raw distributions and computed nanostructure metrics at each $T_5$ are presented in Fig.~\ref{fig:fig_fringe_hists}. The HRTEM images in Fig.~\ref{fig:fig_HRTEM_supp} provide a qualitative overview of the evolution of particle nanostructure with increasing temperature. Visual inspection indicates that the degree of graphitization initially increases from panel (a), developing into a characteristic onion-like structure, and remains broadly similar across panels (b)–(d). At higher temperatures, however, the structures appear more straight (lower tortuosity) and less organized in onion-like shapes. The insets in Fig.~\ref{fig:fig_HRTEM_supp} show the corresponding extracted fringe maps. The overall fringe density and alignment qualitatively reflect the same general trend observed in the images. However, the extracted fringes do not fully reproduce all fringes visible in the original HRTEM micrographs. In this analysis, a single, fixed set of image-processing parameters was applied to all images to maintain methodological consistency while achieving reasonable fringe detection across the dataset. As a consequence, the degree of correspondence between detected and visually identifiable fringes varies among panels. A representative example occurs in panels (c) and (d), which appear to exhibit comparable levels of graphitization by visual inspection. Nevertheless, the extracted fringe maps differ substantially, with panel (d) yielding fewer and more fragmented detected fringes. This discrepancy likely reflects limitations of the automated fringe extraction procedure, particularly in regions containing intersecting fringes or strong intensity variations. Such differences in fringe detection contribute to the divergence between qualitative visual interpretation and the statistical distributions derived from the extracted fringe metrics shown in Fig.~\ref{fig:fig_fringe_hists}.

\begin{figure}[!ht]
    \centering
    \includegraphics[width=1.0\linewidth]{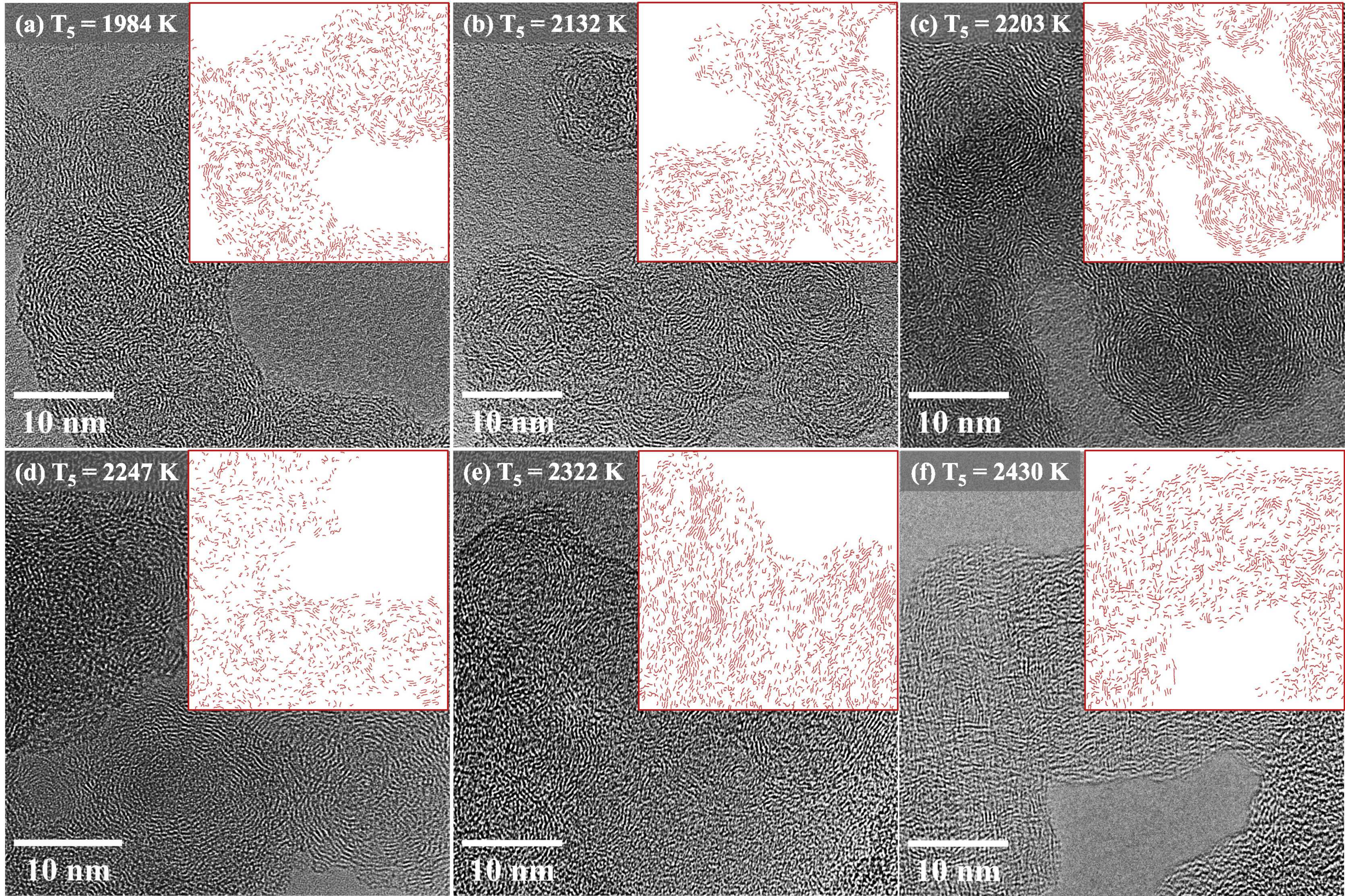}
    \caption{Representative HRTEM images under different conditions with extracted fringes as insets. $T_5$ increases from 1984 to 2430~K across panels (a)–(f). By visual inspection, images in (b), (c), and (d) appear to exhibit close degrees of graphitization.}
    \label{fig:fig_HRTEM_supp}
\end{figure}

\begin{figure}[!ht]
    \centering
    \includegraphics[width=0.5\linewidth]{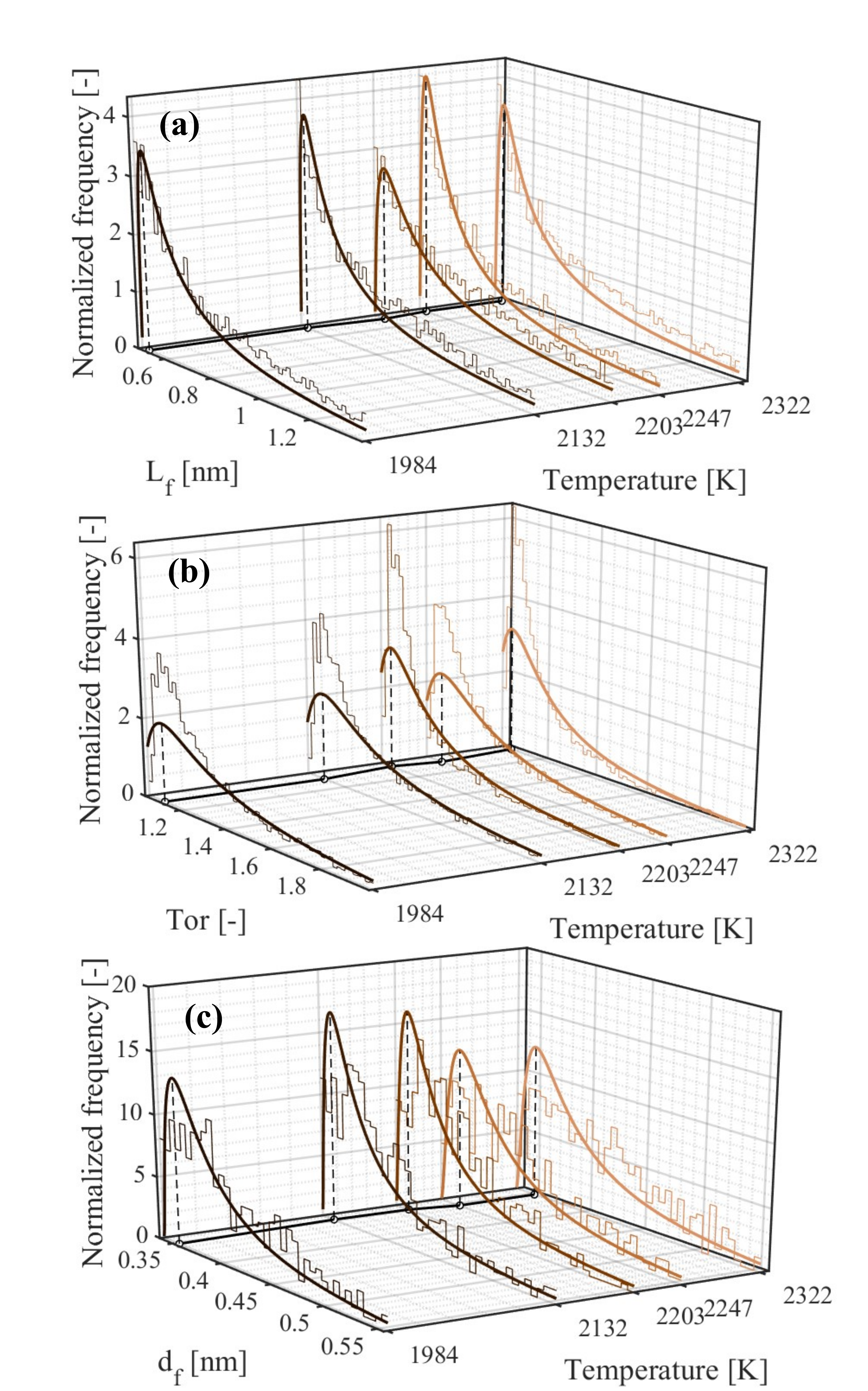}
    \caption{Fringe-statistics distributions for five temperatures: fringe length (a), tortuosity (b), and inter-fringe spacing (c), shown as histograms with shifted lognormal fits. The fitted mode at each temperature is projected onto the base plane (black markers) to present the temperature trend.}
    \label{fig:fig_fringe_hists}
\end{figure}

\FloatBarrier 

\section{Pressure and temperature time histories}
\label{SM:PandT}

\begin{figure}[h]
    \centering
    \includegraphics[width=0.95\linewidth]{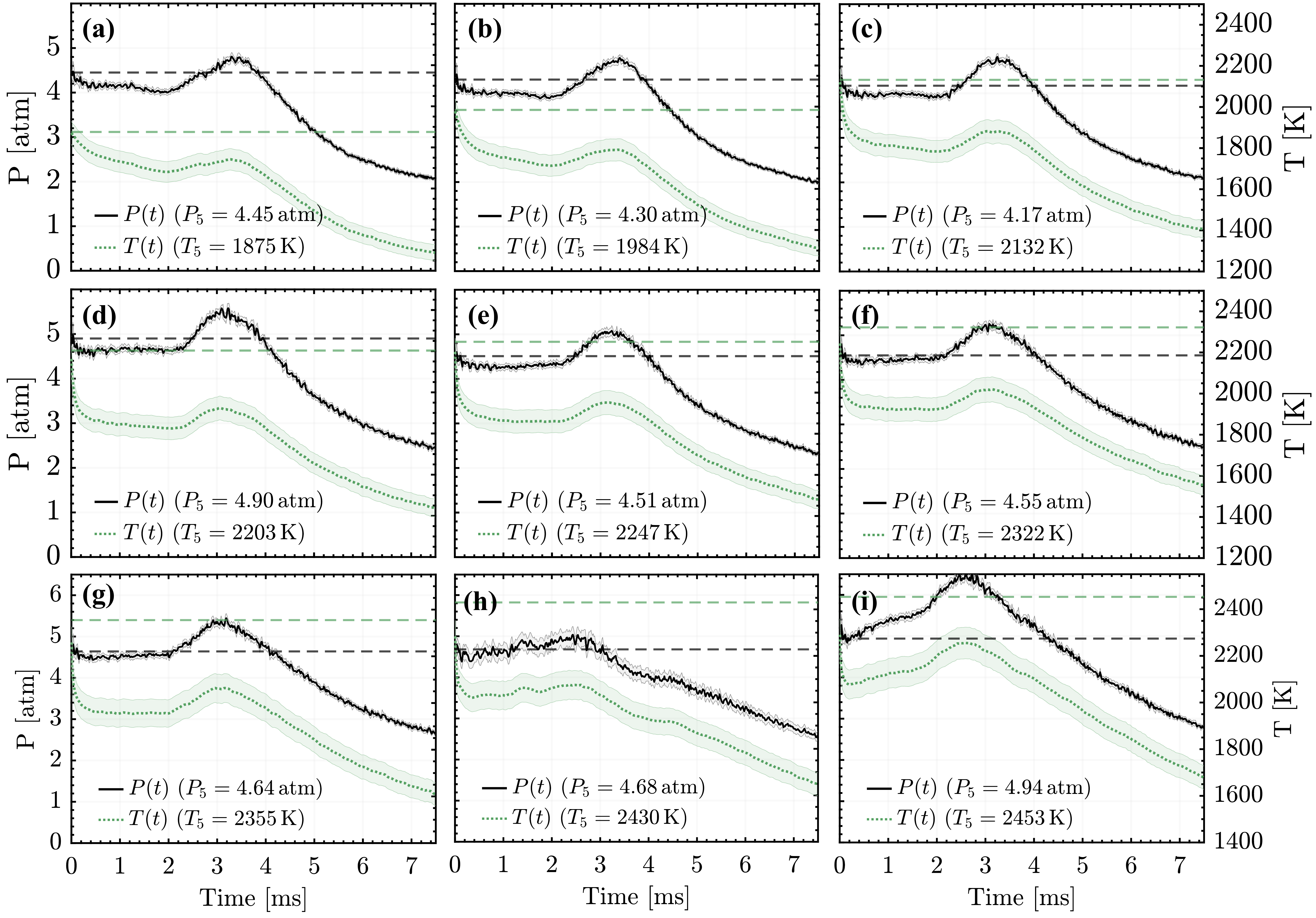}
    \caption{\footnotesize Measured $P(t)$ and modeled $T(t)$ time histories for the reflected-shock \ce{CH4}/Ar pyrolysis experiments considered in this study, ordered from lowest to highest initial post-reflected-shock temperature from panels (a) to (i). Black traces show the measured $P(t)$, recorded using a sidewall-mounted piezoelectric pressure transducer (Kistler 603B1), with horizontal dashed lines indicating the initial post-reflected-shock pressure, $P_5$. Green dotted traces show the corresponding modeled $T(t)$; green shaded bands denote the uncertainty in the inferred temperature history, and horizontal dashed green lines indicate $T_5$.}
    \label{fig:f_supp_PT_hists}
\end{figure}

Figures~\ref{fig:f_supp_PT_hists}a-i show the measured pressure histories and corresponding modeled temperature during all \ce{CH4}/Ar pyrolysis experiments examined here. The filtered experimental pressure trace was imposed directly on the Cantera reactor simulation using the CRECK kinetic model to account for the thermodynamic evolution of the gas after passage of the reflected shock (due to chemistry and gas dynamics). The resulting $T(t)$ profiles show that each experiment undergoes a rapid initial cooling due to endothermic initiation steps, a transient temperature rise associated with the shock reflection off the driver gas contact surface, followed by cooling into the expansion fan. 

\FloatBarrier 

\section{Comparison of induction time metrics}
\label{SM:indTimes}

To measure induction times ($\tau_\mathrm{ind}$) the noise-floor definition was chosen to represent the delay between fuel decomposition and the first measurable extinction signal from condensed-phase formation. By construction, this approach is most sensitive to the onset of detectable particle formation rather than to the subsequent growth rate of the signal. It further enables model comparisons that are nearly agnostic to assumptions of $E(m)$. 

In other work, the inflection-tangent method is commonly used to compare $\tau_\mathrm{ind}$ across studies. This approach is typically less dependent on particular experimental setup and thus works well when particle formation occurs readily, during a quiescent post-shock test time. However, for delayed particle growth, the infleciton tangent approach is more sensitive to processes that control signal growth after inception, including pressure gradients, transient gas-dynamic effects, and other non-ideal shock-tube behavior. This distinction was important for the present \ce{CH4} pyrolysis data, where particle formation is relatively slow. 

\begin{figure}[h]
    \centering
    \includegraphics[width=0.85\linewidth]{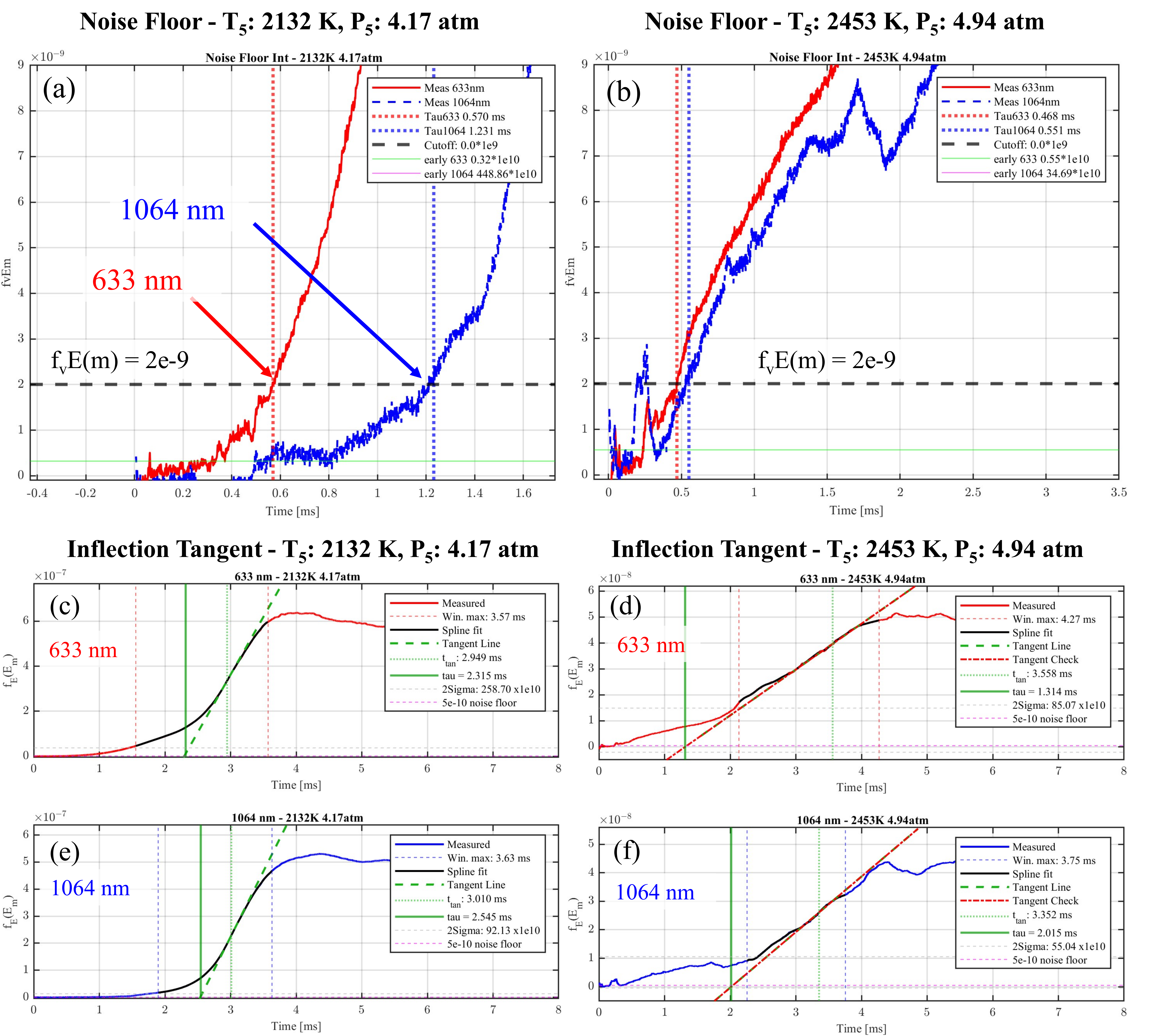}
    \caption{\footnotesize Comparison of two induction-time extraction methods applied to measured 633 and 1064~nm extinction traces. Panels (a,b) show noise-floor crossing times defined at $f_\mathrm{v}E(m)=2\times10^{-9}$, and panels (c--f) show induction times from the inflection-tangent method for mid-temperature ($T_5=2132$~K, $P_5=4.17$~atm) and high-temperature ($T_5=2453$~K, $P_5=4.94$~atm) cases. The noise-floor method is tied to the first measurable extinction signal, while the inflection-tangent method is more sensitive to the later growth behavior of the trace.}
    \label{fig:Supp_indTime_methods}
\end{figure}

Results of noise floor crossing times and the inflection tangent fit to 633 nm and 1064 nm extinction traces from a mid-temperature experiment ($T_5$ = 2132~K, $P_5$ = 4.2 atm) are shown in Figs.~\ref{fig:Supp_indTime_methods}a,c,e, and for very high temperature ($T_5$ = 2453~K, $P_5$ = 4.9 atm) in Figs.~\ref{fig:Supp_indTime_methods}b,d,f. At higher $T_5$, the noise floor approach is limited by low SNR, beam steering, optomechanical vibrations and small variations in laser power, while the inflection tangent approach struggles when growth curves become flatter and less sharply sigmoidal, making the inflection point difficult to define. 

\begin{figure}[h]
    \centering
    \includegraphics[width=0.5\linewidth]{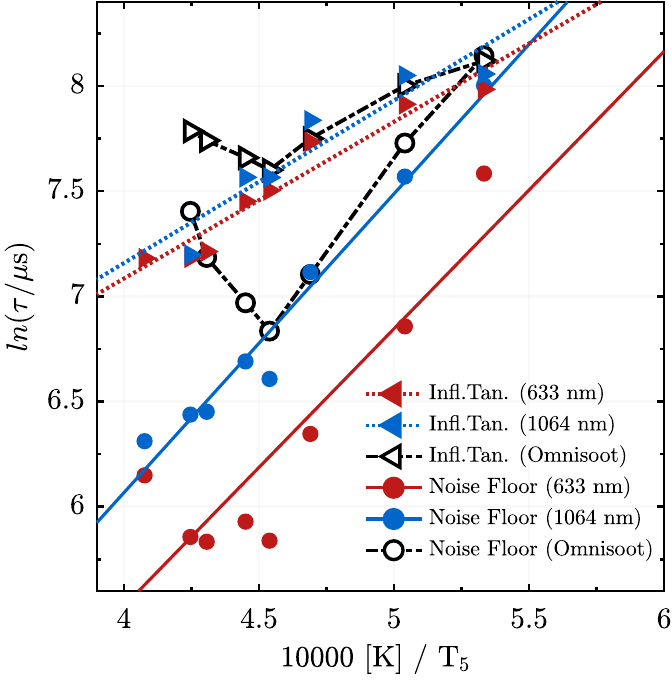}
    \caption{\footnotesize Induction times extracted by the noise-floor (circles) and inflection-tangent (triangles) methods for measured 633 (red) and 1064~nm (blue) extinction signals and for Omnisoot predictions (white fill), shown in Arrhenius form. Both methods give similar $\tau_\mathrm{ind}$ at lower $T_5$, but the inflection-tangent method returns slower induction times as $T_5$ increases. The relative delay between 633 and 1064~nm is preserved with either method, and Omnisoot shows the same low-temperature agreement with 1064~nm results but a non-Arrhenius delay above approximately 2200~K.}
    \label{fig:fig_indTimes_nf_infl_comp}
\end{figure}

Induction times ($\tau_\mathrm{ind}$) using the noise-floor and inflection-tangent methods for both the multi wavelength extinction data and for Omnisoot predictions are shown in Arrhenius format in Fig.~\ref{fig:fig_indTimes_nf_infl_comp}. Inflection tangent results clearly return similar $\tau_\mathrm{ind}$ to the noise-floor at lower $T_5$, but much slower $\tau_\mathrm{ind}$ as $T_5$ increases. Using either approach, however, $\tau_\mathrm{ind}$ exhibits similar temperature-dependent delays as wavelength increases from 633 to 1064 nm, and the results of Omnisoot predictions indicate the same low-temperature match with 1064 nm results, but non-Arrhenius delay at temperatures above 2200 K.

\FloatBarrier 


\newpage

\small
\baselineskip 10pt

\end{document}